# Research on discharges in micropattern and small gap gaseous detectors

V. Peskov[1,2], P. Fonte[3]

[1]ICN UNAM, Mexico
[2] CERN, Geneva
[3]ISEC and LIP, Coimbra Portugal

**Abstract**

This report summarizes the present knowledge on discharges in micropattern and small gap gaseous detectors and the physical mechanisms involved. These include the space-charge (Raether's) limit, rate-induced breakdown, cathode excitation effect and electron emission from the cathodes in the form of jets, inter-GEM breakdown in multistep configurations and finally surface streamers.

# 1. Introduction

Classical large-gap gaseous detectors: single wire-counters, MWPCs, large-gap (1-3 cm) parallel-plate chambers –usually are very robust and cannot be destroyed by spurious discharges appearing during their operation. Moreover they are operated in modes which are safe for the given optical or electronic signal detecting system: pulsed HV in the case of the parallel-plate detectors, Geiger mode or limited streamer mode in the case of the wire detectors.The physics of the discharges in this type of the detectors is well understood today (see recent review talks [1-3]).

In contrast, most of the recently introduced gaseous detectors: micropattern and small gap gaseous detectors (MP/SGD)- are quite fragile and can be easily damaged by discharges. Thus the study of the physics of the discharges in this new type of detectors and searching the ways of MP/SGD protection against harmful discharges is a very important practical topic. One should add to this that in many applications (high luminosity colliders, medical applications and so on) MP/SGDs should operate at extreme conditions when the spurious discharges can appear with increased probability, for example at very in high counting rates or in presence of heavily ionizing particles, or at exceptionally high gas gains (detection of single electrons).

This report summarizes our studies of feature of discharges in these typed of gaseous detectors which we carried with various collaborators for the last 10 years (see appropriate references through out the text) As follows from these studies there are several key issues to be addresses:

1. Raether limit
2. Feedbacks
3. Rate effect
4. Cathode excitation effect
5. Discharge propagation in cascaded detectors
6. Surface streamers

It will be useful to consider first discharge phenomena separately at low and high counting rate and then consider some more "exotic" discharge mechanisms such as jets, a cathode exaction effect, breakdown propagation from on detector to another (when they operate in cascade), surface streamers an so on.

# 2. Low counting rates

## 2.1. Raether limit

In the case of the poor quality detectors, the discharges are triggered by the presence of microdefects: sharp edges, micro-particles remaining after the production both inside and outside the holes, dirty spots (which are often semiconductive) and so on. However, it is quiet well establishes today [4] that in the case of good quality MP/SGD, the breakdowns appear when the total charge in the avalanche reaches some critical value:

$$Q_{crit}=A_{max}n_0 \sim 10^6\text{-}10^7 \text{electrons (1.1)},$$

where $A_m$ is the maximum achievable gas gain and $n_0$ is the number of primary electrons created by the radiation in the active gas volume of the detector. This limit is very similar to those which was established a long time ago by H. Raether for large gap (0,5-5 cm) parallel-late chambers [5] and this is why we also called it a Rather limit. As follows from formula (1), in the case of the detection of single electrons ($n_0$=1) the $A_{max}$ can be as high as $10^6$. However, in the case of the detection of radiations producing $n_0 \gg 1$ primary electrons the maximum achievable gain will be accordingly reduced. For example, in the case of the detection of x-rays form a $^{55}$Fe radioactive source (each photon creates $n_0 \sim 220$ primary electrons), the maximum achievable gain will be $\sim 10^4$ and in the case of alpha particles ($n_0 \sim 10^5$ electrons) the maximum sustainable gains will be below $\sim 10^2$.

Hence, if one uses MP/SGD for the detection of single photoelectrons (so the gas gain should be high, about $10^5$-$10^6$), any radioactive background creating $n_0$>1 primary electrons will trigger breakdowns. Therefore unfortunately, sparks are almost unavoidable at high gain operations.

The exact value of $Q_{crit}$ depends on several factors, the most important among them are: geometry and density of the primary electron cloud, value of $n_0$ [4], detector geometry and electron and ion diffusion in the given gas and the given electric filed [6]. By optimizing these parameters one can in principle increase $Q_{crit}$ and therefore $A_{max}$. As an example **Figs. 1** and **2** show gain curved measures with UV and x-rays for TGEM operating in Ar+5%CH$_4$ and in pure Ne at 1 atm. As one can see in Ar+5%CH$_4$ the maximum achievable gain measured with X-rays is 100 times less than with single photoelectrons (UV) in a good agreement with the Raether limit. In Ne however, this difference reduces on a factor of 10 which can be explained by the lower operating voltage and larger photoelectron tracks in this gas (see **Fig. 3**)

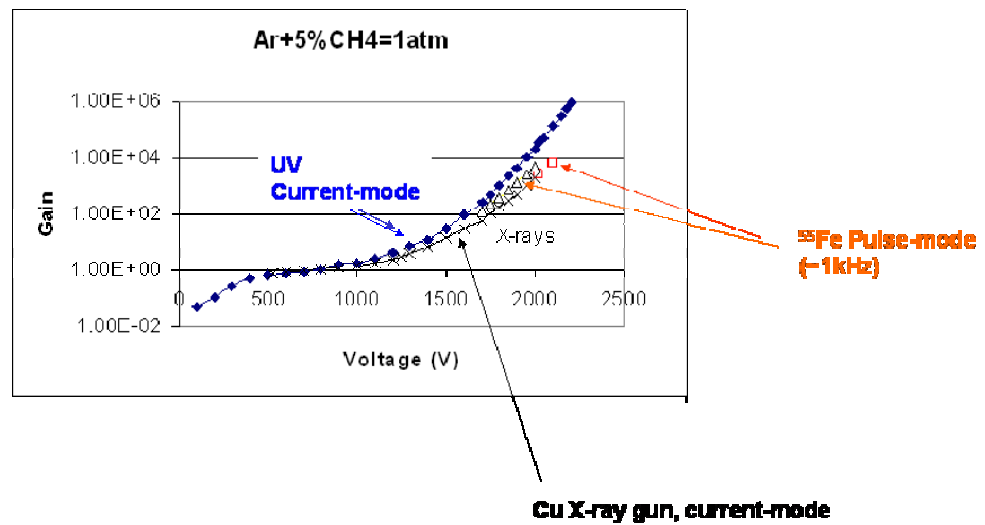

Fig. 1. Gain vs. voltage curves measured in Ar+5%CH4 with UV and 6-9keV X-rays with a single TGEM (TGEM geometry: holes diameter 0.5mm, pitch 1mm, thickness 0.8mm, rims 0.1mm ). The measurements were stopped when the first breakdown appeared (from [7]).

One can see that in this gas mixture the maximum achievable gain with X-rays is almost on two orders of magnitude lower than in the case of detection single photons only as one can expect from the Raether limit (see eq. (1.1)).

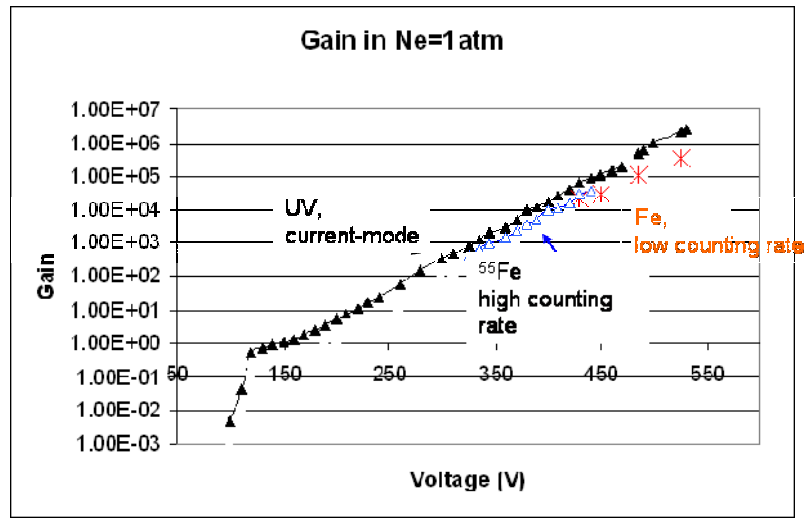

**Fig. 2.** The same measurements, but performed in Ne [7]. As one can see, in Ne the difference in the maximum achievable gain is only about a factor of ten (see explanations in the text)

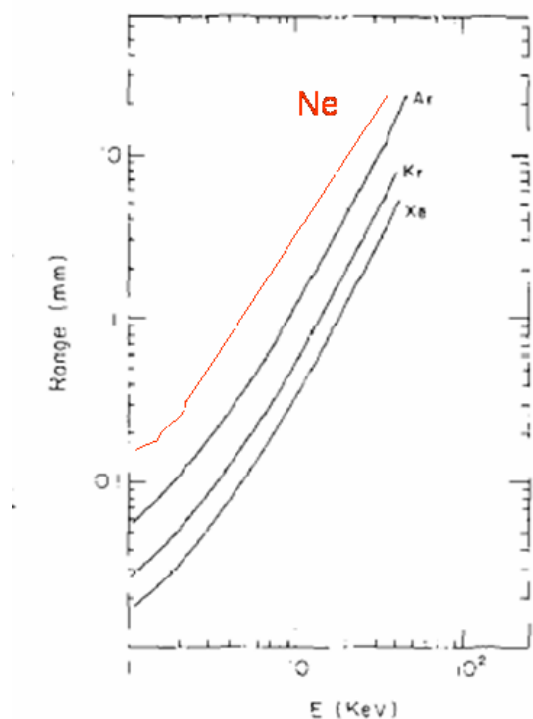

**Fig. 3.** Calculated mean length of photoelectron tracks in Ne and other noble gases. As can be seen the mean free pas of photoelectrons produced by Fe is about 1mm. This is ~5 times larger than in Ar which leads to lower density of ionization and higher Raether limit (which depends on the density of primary ionization)

The Raether limit explains well why the maximum achievable gain of almost all type of MP/SGD s measured with 6 keV is always ~$10^4$.

It also explains why the maximum achievable for all MP/SGD s gain drops with pressure (see for example **Fig. 4** reproduced from [4])

Because the value of $Q_{crit}$ depends on the size of the discharge gap and on the density of the primary electron cloud there are rooms for the optimization of these parameters in order to reach the highest possible value of the $Q_{crit}$. For example, as was shown in [6] in double or triple GEMs there is (due to the electric field geometry), an enhanced diffusion of electrons when they are extracted from the GEM holes and as a result the cloud of electrons is noticeably expanded so that with the double and the triple GEMs higher overall gains can be achieved than with the single one.

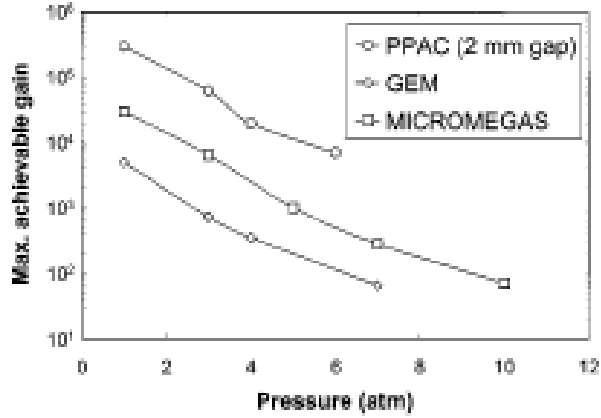

**Fig. 4.** Typical dependence of $A_{max}$ vs. gas pressure [4].

## 2.2. Feedbacks

When the (MP/SGD) are operating in poorly quenched gases or combined with photocathodes, the maximum achievable gain $A_{mf}$ can be additionally restricted by the feedback mechanism (see [8] for details) so that

$$A_{mf}\gamma k = 1 \quad (I.2),$$

where k is a coefficient determining what fraction of ions (in the case of the ion feedback) or photons (in the case of the photoeffect) reaches the cathode of the hole-type detector (in the case of wire-type or parallel-type detector usually k~1) and

$\gamma$ is the probability of secondary effects (which depends on the gas and on the electric field E on the cathode surface and [8, 9]). As it is well know, the feedback loop can be caused by photons or by ion recombination on the cathode (with secondary process coefficients $\gamma_{ph}$ and $\gamma_+$ respectively) as well as by the combination of these processes [8]. As was follows from [8, 9]:

$$\gamma_+ = b(E)(\varepsilon_i - 2\varphi) \quad (1.3)$$

$$\gamma_{ph} = \int Q(E, E_v) S(E_v) \, dv \quad (1.4),$$

where b is a coefficient, $\varepsilon_i$ is an ionization potential of the gas, $\varphi$ is the cathode work function, $E_v$ is the photon energy, $S(E_v)$ is the avalanche emission spectrum in the given gas and $Q(E, E_v)$ is the quantum efficiency of the cathode in the given gas mixture.

As can be seen from formulas (1.3) and 1.4) as well as from **Fig. 5**, $\gamma_+$ is linearly increasing with the ionization potential of the gas and $\gamma_{ph}$ is very sharply increasing with the photon energy $E_v$ (mostly because Q(E, Ev) sharply increases with $E_v$).

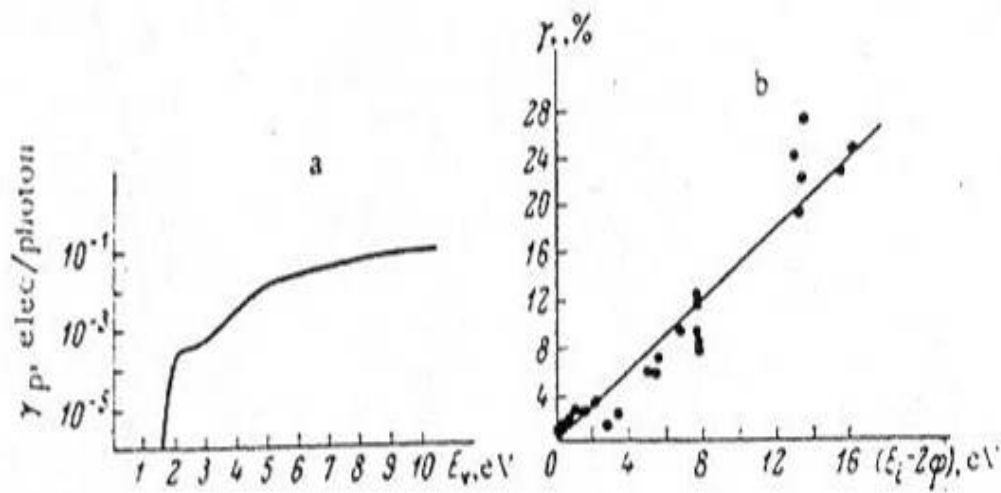

**Fig. 5.** a) dependence of $\gamma_{ph}$ on photons energy $E_v$ for CuI photocathode [9], b) plot of $\gamma_+$ vs. $\varepsilon_i-2\varphi$, where $\gamma_+$ is probability for ion extract an electron from the cathode, $\varepsilon_i$-gas ionization potential, $\varphi$- work function of the cathode [9].

Usually appearance of the photon feedback declares as gain curve deviation from linear behaviour in log. scale (see **Fig. 6** as an example). This is because in the case of the MP/SGD s the photon feedback pulses appear with a time delay $\tau-\ll T_{int}$, $T_{dif}$, where $T_{int}$ and $T_{dif}$ are the integration and differential time constants of the shaping amplifier (note that to achieve the best ratio signal to noise usually one have to keep $T_{int}=T_{dif}$). In some designs of MP/SGD s ion feedback pulses also may have a rather short time delays compared to the primary pulse ($\tau+\ll T_{int}$, $T_{dif}$) and this also causes the gain curve deviation from the straight line in logarithmic scale. However, usually $\tau+\geq T_{int}$, $T_{dif}$, so that one can clearly observe these pulses as well as the increase of the counting rate with the gain.

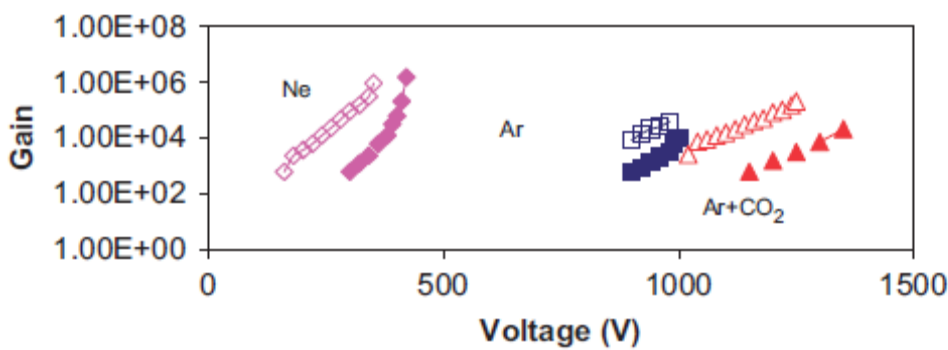

**Fig. 6**. Gain vs. voltage curved measured with RETGEM in various gases. One can see that in the case of Ne and Ar at high gas gains the gain vs. voltage curves start deviate from straight line due to the feedback mechanism (from [10])

Note that very often, particularly in the case of photocathodes sensitive to visible light $A_{mf}<A_{mr}$ [11].

In practice, because in GEM detector the cathode is geometrically shielded from the direct light emitted by the avalanches, the GEM experience only ion feedback and can operate at relatively high gains in badly quenched gases including noble gases. However, cleaner is the noble gas, the lower is $A_{mf}$ [12]. In ultraclean He and Ne practically no gas gain was achieved with the GEM detectors [13]. This was already clearly observed in the case of wire–type detectors [9] and is well understood. In very clean noble gases the mean free pass of noble gas ions before they experience a charge transfer with the molecules of impurities increases, so at some level of cleanness the majority of them can reach the cathode. Because the $\gamma_+ \sim \varepsilon$, the ions of noble gases, due to their higher ionization potential then the ionization potential of molecules of impurities, are capable to trigger a feedback loop with a consequent breakdown.

Detectors the cathode of which are not geometrically shielded from the avalanche emission, for example MICROMEGAS or PPAC can operates at high gains only in quenched gases; in noble gases due to photon[1] and ion feedbacks only very low gains are possible to achieve.

Recently an interesting effect was observed [12]: in the case of TGEM combined with a CsI photocathode and operating in Ne the breakdown at extremely low UV fluxes occurs via a streamer mechanism (when the Raether limit was reached) whereas as high fluxes, due to the cathode excitation effect (see next paragraph) the breakdown starts occurring via the feedback mechanism. So in some case the breakdown phenomena may depends on the intensity of the UV flux

## 3. High counting rates

### 3.1. Features of breakdowns at high counting rats

It is also quite well established today that the maximum achievable gain for all MP/SGD s drops with the counting rate [15]. As an example, **Fig. 7** shows the measurements of $A_{max}$ performed with 6keV photons for various micropattern gaseous detectors. The dark area in this figure is the region where breakdowns appear. One can replote this sparking "boarder" in a more general way: $Q_{max}$ vs. rate as was it done in Fig**. 8.**

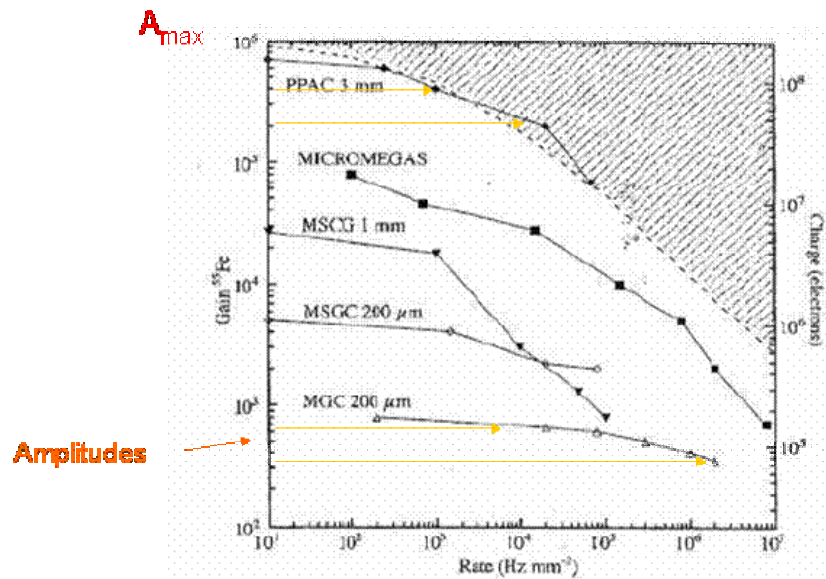

Fig. 7. Maximum achievable gain vs. rate plotted for several detectors. Measurements of $A_{max}$ were performed with 6 keV photons [15].

---

[1] Photon feedback in noble gases usually is very strong. This is because Q (E, $E_v$) sharply increases with photon energy and also the excimer emission spectra of noble gases S (Ev) are in extreme ultraviolet [14].

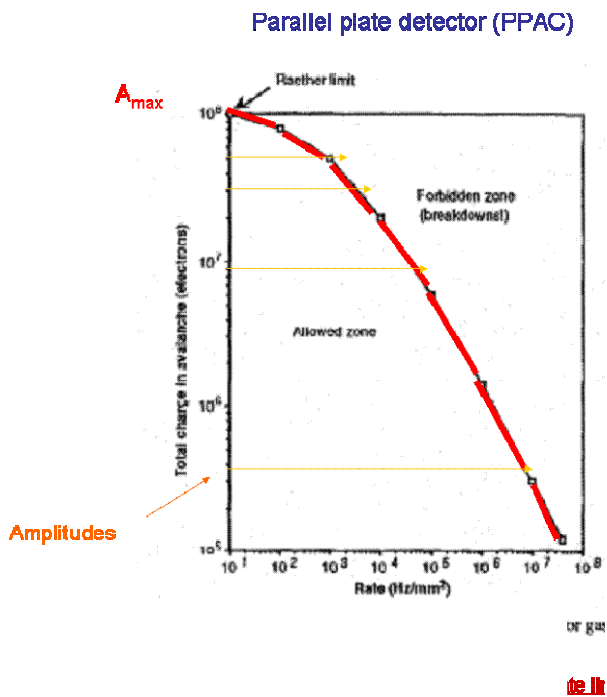

**Fig. 8**. $Q_{max}$ vs. rate for gaseous detectors. Note that usually the signal amplitude does not drop with rate, however there is a rate limit for each amplitude [16].

## 3.2. Physical mechanisms involved in breakdowns at high counting rates

What is the physical mechanism behind this interesting and in fact very fundamental phenomena- a maximum achievable gain drop with rate?

As follows from earlier studied two mechanisms contribute to the rate effect (see for example [2]): statistical avalanche overlapping and electron jets.

### *3.2.1. Statistical avalanche overlapping*

One may ask whether the observed decrease in the maximum spark-free gain at higher rates may be the consequence of the merging of several avalanches, adding the respective electrical charge to form a space-charge field comparable to the applied field and triggering the formation of a streamer. The probability of such an event obviously increases with the counting rate.

Physically, this requires that a sufficient number of avalanches will overlap in time and space within some effective distance and time interval – a "superimposition cell" (SC). The detailed determination of the dimensions of a SC is a complex matter that can only be tackled by 3D avalanche calculations. In here we will take some educated guesses, considering such dimensions to be comparable to the multiplication region (gas gap, GEM hole, etc) and to the ions transit time.

If one depicts a 1 second long frame of the impinging particle beam with cross-sectional area $A$ and rate density $R$ as points lying within a cylinder of "volume" $V = A \times (1s)$, as shown in **Fig. 9**, the probability $P(n)$ of finding $n$ avalanches within a SC of "volume" $v = a\tau$ is given by the Poisson distribution with parameter (mean value) $\lambda = Rv$. There are $N = V/v$ superimposition cells in a time frame.

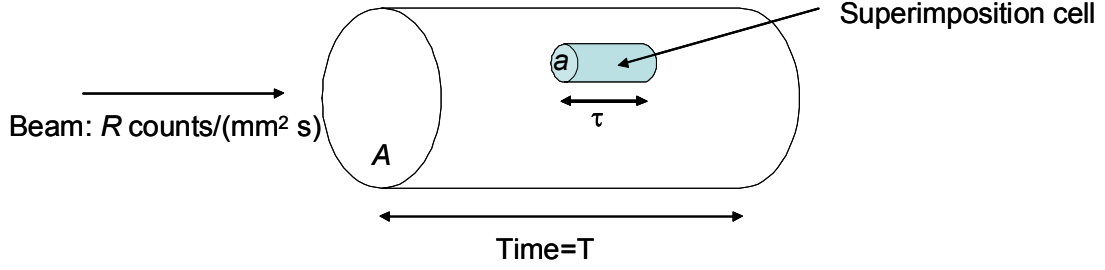

**Fig. 9.** A schematic illustration of the calculation model

Let's denote by $p$ the probability of sparking in any SC. This is the probability that the total charge present within a SC, $nq$, where $q$ is a typical avalanche charge that depends on the gas gain, will exceed the space-charge limit $Q_d$: $p = P(n > Q_d/q, \lambda) = 1 - T(Q_d/q, \lambda)$. In here we have denoted by $T(x, \lambda)$ the cumulative Poisson distribution with parameter $\lambda$.

We denoted the space charge limit as $Q_d$ as a reminder that it should not be an universal constant, but should depend on each specific type of gas amplification structure. For instance, for planar gas gaps more than a few millimetres wide, $Q_d$ will be close to the classical Raether's limit of $10^8$ electrons. However, for other geometries, most notably micropattern detectors that develop avalanches within very tiny structures, $Q_d$ may be smaller by several orders of magnitude, since, roughly, the electric field at the surface of a sphere containing a fixed amount of charge depends quadratically on the inverse of its radius.

In practical terms, to measure the spark probability, $S$, on a detector the conditions should be chosen such that the absolute spark rate will be not so large that the detector or it's electronics doesn't operate at all or so small that beam-induced sparks are almost never observed, overshadowed by more frequent highly-ionizing background events, such as air showers or $\alpha$ decays from airborne radioactivity or from chamber materials. A reasonable round figure might be about once per one hundred seconds: $P(spark) = S \sim 1/100\, s^{-1}$.

The probability of sparking is the complement of the probability of not sparking in any of the $N$ SCs: $S = 1 - P(not\, spark) = 1 - (1 - p)^N$, which for very large $N$ and small $S$ turns into an identity if $p = S/N$.

The maximum individual avalanche charge, $\tilde{q}$, is then given by the solution of $p = 1 - T(Q_d/\tilde{q}, \lambda)$, or

$$\frac{1}{\tilde{n}} = \frac{\tilde{q}(\lambda)}{Q_d} = \frac{1}{T^{-1}(1-p, \lambda)} \quad (2.1)$$

where $T^{-1}$ is the Poisson inverse cumulative distribution function, also known as the percentile function. As $T^{-1}(x, 0) = 1 \Rightarrow \tilde{q}(0) = Q_d$, the expression defines the rate induced admissible-gain drop factor. This is the same as the inverse of the number of avalanches required for sparking in one cell, $\tilde{n}$. The function is depicted in **Fig. 10**

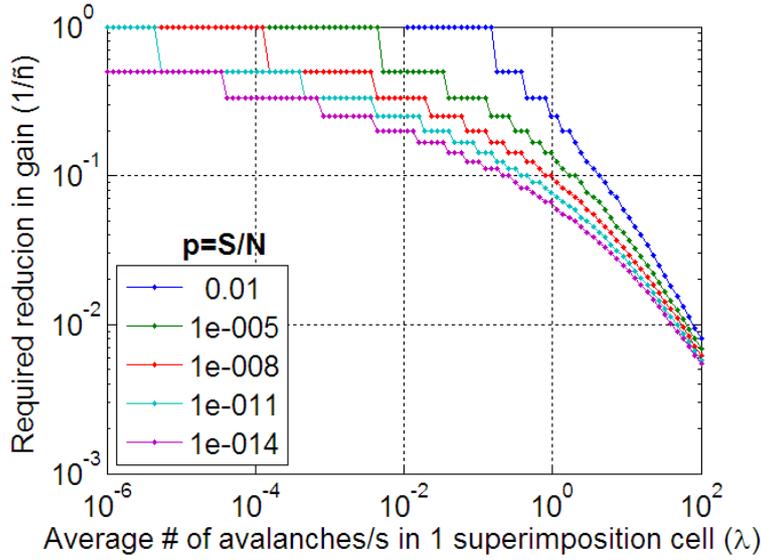

Fig. 10. The rate-induced gain drop factor defined in eq. 2.1), as a function of $\lambda = R a \tau$ and $p = S/N$. The function is quite insensitive to the value of $p$.

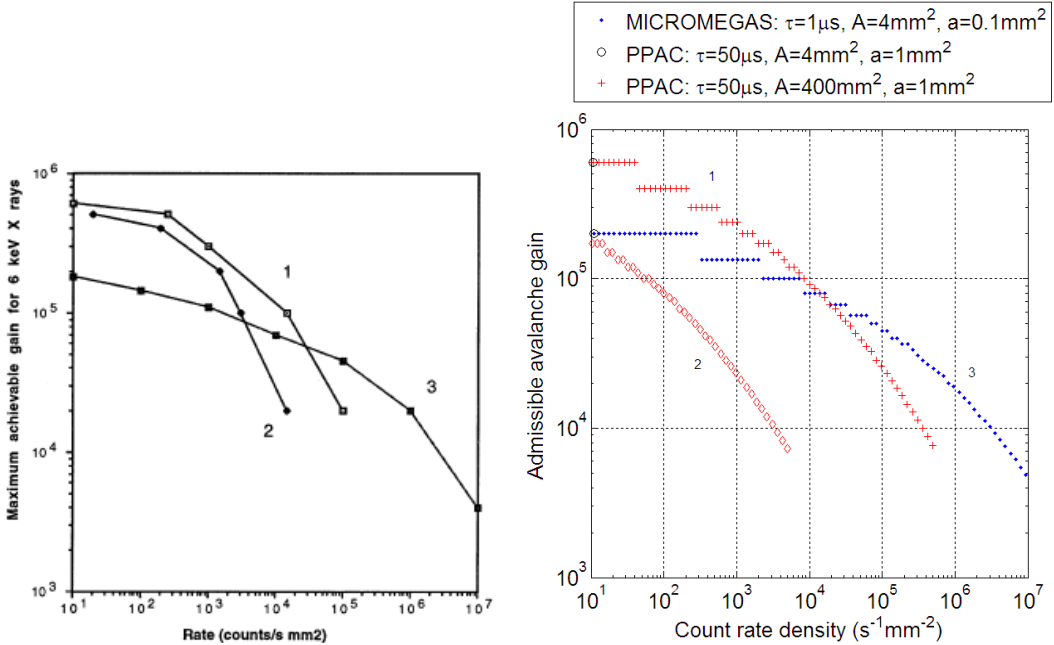

Fig. 11. Comparison between experimental data on rate-induced breakdown in parallel geometry detectors, PPAC and MICROMEGAS [15], and eq.2.1. The calculation was performed using the estimated model parameters indicated and the value of $Q_d$ for both detectors was adjusted so that the two encircled points correspond to the experimental data. Although the model seems to overestimate the influence of the beam cross-section, the resemblance is inescapable.

In **Fig 11** it is shown a comparison between experimental data on rate-induced breakdown in parallel geometry detectors and eq.2.1. Although the model seems to overestimate the influence of the beam cross-section in PPAC, the resemblance is inescapable.

It is cleat that, with a much narrower gap, the SC of MICROMEGAS is much smaller than PPAC's, so MICROMEGAS's maximum gain is much less affected by the count rate. However, this comes at the penalty that the narrower gap features a lower space-charge limit (denser avalanche), so the low rate gain of MICROMEGAS is smaller than PPAC's up to a rate density of about $10^4$ $s^{-1}$ $mm^{-1}$.

Another interesting observation is that the curves are never flat, even at very low rates. This is because when the detectors operate close to their space-charge limit, any superimposition will trigger a spark. Even if is this is very infrequent at low rates, the probability is never completely negligible.

One may also wonder whether the avalanche "jets" described in the next section may be just due to the statistical time (and space) overlap of avalanches that triggers the spark itself. As such jets correspond to instantaneous count rates much larger than the average count rate, one should compare both quantities. For instance, taking the last point of curve 1 in **Fig.11,** it corresponds to roughly 100 avalanches (that is, $ñ$) in the 50 μs long SC, corresponding to an instantaneous count rate of 2 MHz. For the same point the average count rate will be A×R=4×($5\times10^5$)=2MHz. Therefore, the statistical accident that created the spark will increase the instantaneous count rate by about a factor 2, which would be hardly noticeable in an oscillogram. Roughly the same conclusion can be extracted for the last point of curve 3 and the effect is even smaller for curve 2. Therefore, the observed avalanche "jets" are unlikely be just statistical accidents, besides the fact that statistics will never explain the observed pre-spark sustained current growth for a fraction of second before the actual spark.

### *3.2.2. New recently discovered phenomena involved in breakdowns at high counting rates: cathode excitation effect and electron jets*

Besides statistical avalanche overlapping effect, certainly there are another phenomena (by the way very exotic!) which are involved in the high rate breakdown. It was observed experimentally that a so called cathode excitation effect and/or jets also significantly contribute to the high rate breakdowns. As an example **Fig. 12** shows avalanche current measured in PPAC irradiated by x-ray gun. One can see that just before the breakdown (sometimes even a 1sec before the breakdown!) the current starts spontaneously rising. Part of this effect can be attributed to the so called a cathode excitation effect (see next paragraph for more details). However, very often there is as well another type of "preparation "process: spontaneous pulses of very high amplitudes (we called them jets) measured on the top of the discharge current-see **Fig .10.** so that the slow current rice in **Fig. 12** is due to the integrated large number unresolved pulses. The existence of two mechanisms- the cathode excitation and jets is clearly demonstrated by -see **Fig.14**, when the same measurements were performed with much better time resolution ~20 ns comparable with the duration of spontaneous pulses.

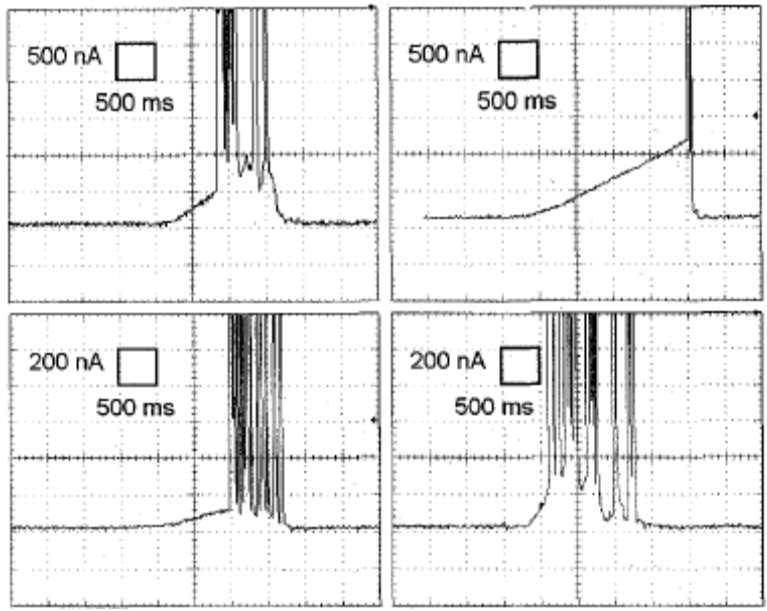

**Fig. 12.** Current measured from a PPAC irradiated by x-ray gun at gas gains close to breakdown [17]. One can see that before the breakdown discharge current starts spontaneously increases till the detector transit to a discharge. In most cases the current rise is due to the integrated large number of pulses shown in **Fig. 13**, where the measurements where performed with much better time resolution

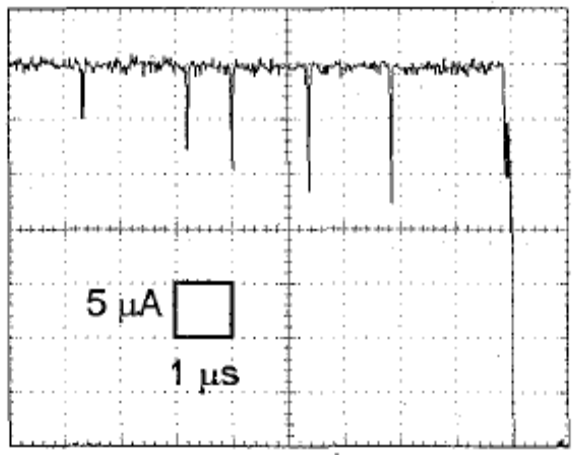

**Fig. 13.** Detail of the gap current (see **Fig. 11**) just before the breakdown [17]

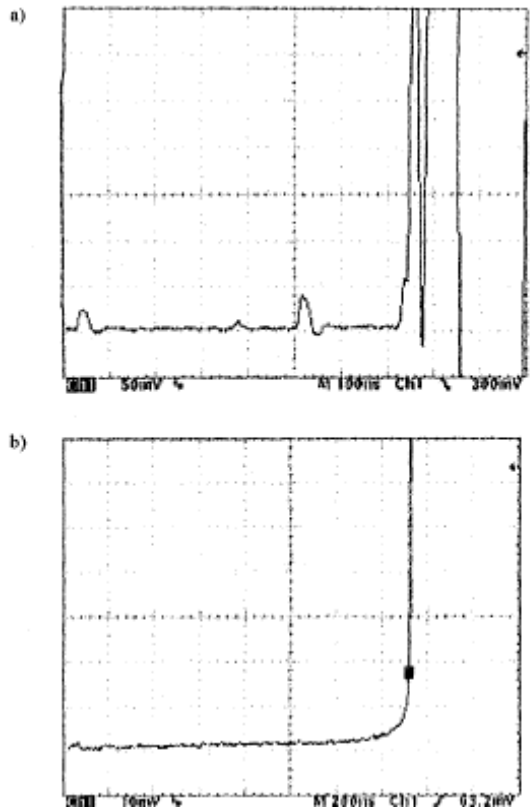

**Fig.14.** Two types of pre-breakdown phenomena recorded with 20ns time resolution: a) pulses on the top of the constant discharge current, b) steady current rise started 100ns before the breakdown [16]

At very high rates (>$10^7$Hz/mm$_2$), additionally to the cathode excitation effect, plasma-type effects may appear as well. They include the modification of electrical field in the cathode region due to the steady space charge, multistep ionization, gas heating effect and accumulation of excited atoms and molecules [18]. As it was shown in other studies, these mechanisms may create instability leading to breakdown [19, 20].

Note that the described above breakdown mode, having in advance of the breakdown a "preparation" activity in form of spontaneous slow current grow or via high amplitude pulses, is not mentioned in known to us literature and can be classified as a completely new phenomena.

Below we will present what is known about this effect up to now

## 4. Further details on cathode "excitation" effect

### 4.1. Metallic cathodes

The cathode excitation effect is a very interesting physic phenomenon. In "macroscopic" way it manifests itself as a hysteresis: after breakdown one cannot immediately apply the same voltage as it was before the breakdown ($V_{max}$), for some time the detector accepts ($V_{ac}$) only lower voltages see **Fig. 15**. Let's analyze this figure. For simplicity lets consider a single-wire counter in which this effect was clearly observed (measurements were done with Cu and stainless steal cathodes).

Suppose that 1 min after the discharge one applies to the single-wire counter a voltage $V_{ap} > V_{ac}$, then a corona discharge immediately will be ignited.

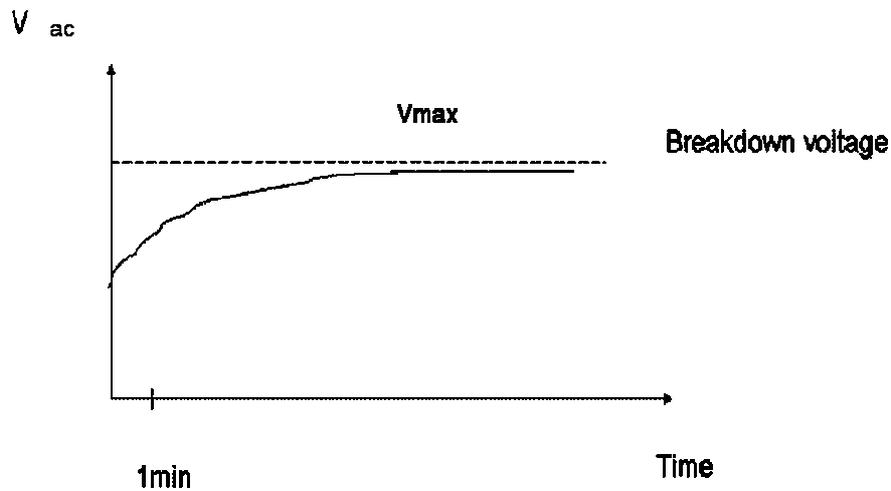

**Fig. 15.** Maximum voltage which detector accepts ($V_{ac}$) vs. time (the breakdown occurred at time=0) measured with a single-wire counter. A dash line indicates $V_{max}$-the voltage at which breakdown occurs at time=0.

Note that similar curves (but with different time scale) were observed for most of gaseous detectors.

The conditions of the corona discharge is

$$A\gamma_{ph}=1 \quad (3.1)$$

or

$$A\gamma_{+}=1 \quad (3.2),$$

where $\gamma_{ph}$ and $\gamma_{+}$ are the probability for the avalanche to create secondary electrons due to the photoeffect or by neutralization of ions on the cathode. Because the gas gain exponentially depends on the applied voltage, at $V_{ap}$ the gas gain is considerably less than at $V_{max}$ and thus $\gamma_{ph}$ and/or $\gamma_{+}$ are temporally (often for minutes!) increased. This conclusion was verified experimentally. **Fig. 16** shows one of the apparatuses used in the past for measuring $\gamma_{+}$ and $\gamma_{ph}$ immediately after the discharge was terminated [21].

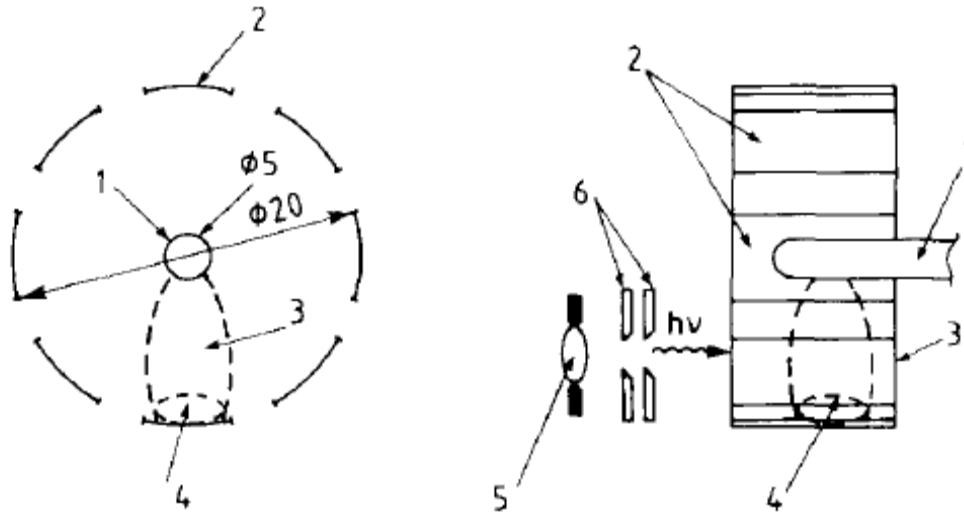

**Fig. 16.** A detector with segmented cathode for measuring $\gamma_+$ after switching a glow discharge. 1-cylindrical anode, 2-segmented cylindrical cathode, 3-glow discharge, 4-cathode spot of the glow discharge, 5-external UV source, 6-collomator (from [21]).

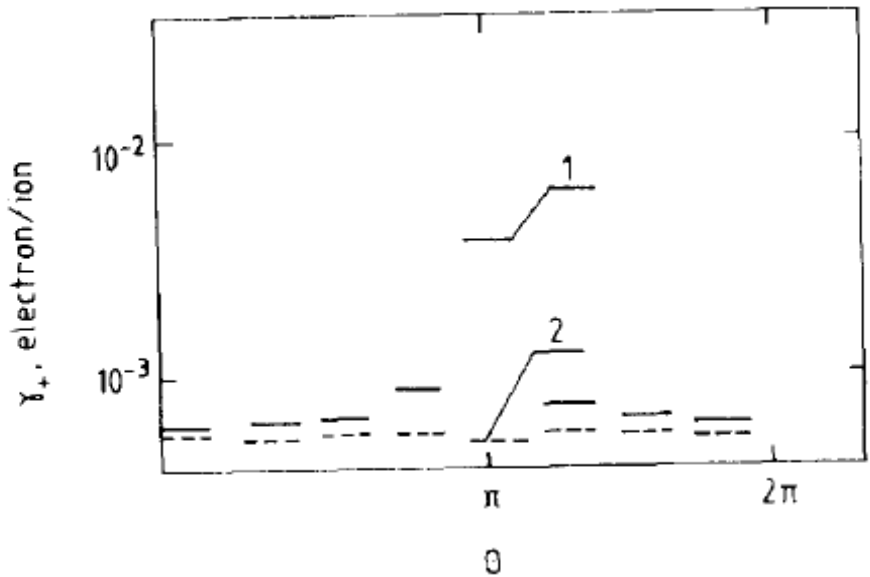

**Fig. 17**. Values of $\gamma+$ as a function of the angle $\theta$ between the upper section(see **Fig 16**) and the investigated one for different time delays $\tau$ after ubrupting the glow discharge: 1) $\tau=9\mu s$, 2) $\tau=10ms$ (from [21]).

By readout different cathode sections of this device (exposed and not exposed to the discharges), it was demonstrated that the cathode area exposed to the discharge had for some time interval enhanced values of $\gamma_+$ and $\gamma_{ph}$-see **Fig 17**.

There were also recent measurements (performed by the authors of this report)demonstrating that after the terminating the corona discharge the sensitivity of the metallic cathode to the UV and visible light temporally increases and very often this effect also accompanates with temporal increase in rate of spurious pulses. Some typical results are presented in **Fig 18 - 20.**

Note that $\gamma_{ph}$ is usually increases with the increase of the cathode quantum efficiency so one can speculate that **Fig. 18** indirectly indicates that after the corona discharge $\gamma_{ph}$ temporally increased.

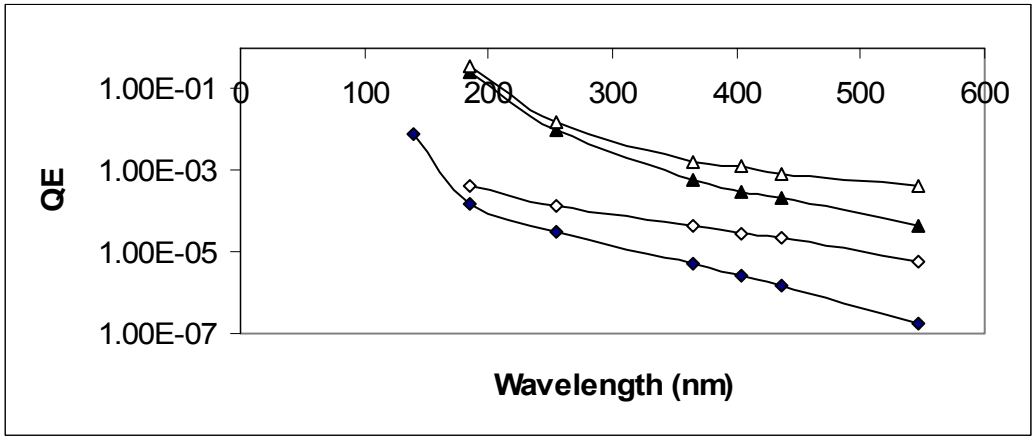

**Fig. 18**. Quantum efficiency vs. wavelength of metal (rhombus) and CsI (triangles) cathodes measured in as ingle-wire counter before a corona discharge (solid symbols) and immediately after the corona discharge was interrupted

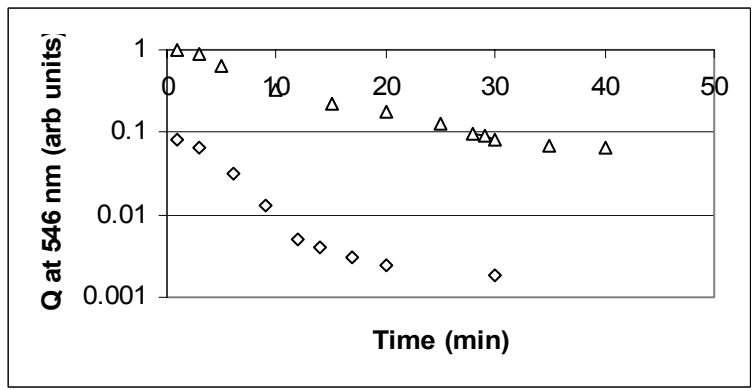

**Fig. 19**. Changes in QE vs. time for Cu (rhombus) and CsI (triangles) photocathodes

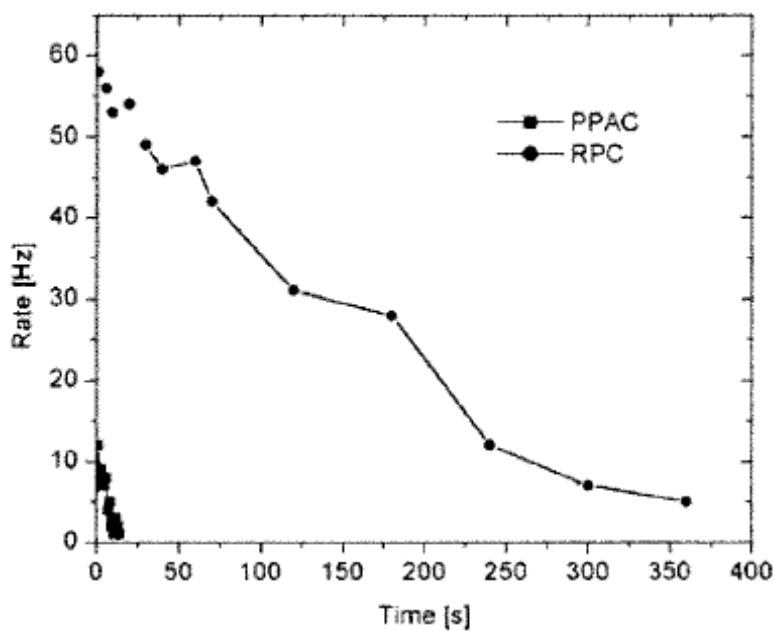

**Fig. 20**. Rate of spurious pulses vs. time measured from a parallel-plate avalanche chamber (Cu electrodes) and from an RPC (Si cathode) operating at the same gas gain in avalanche mode. Gas mixture Xe(20%)+ Kr (40%)+ $CO_2$(20%) at 1 atm [22].

The explanation of all effects mentioned above is a formation of temporal positive ion layer on the on the cathode surface which due to the creation of strong local electric field reduces the work function and also may causes electrons jets (see more details in paragraph 4 and [16]).

This phenomenon is directly related to a well known Malter effect, however strictly speaking a term "Malter effect" is usually used to explain a single electron after-emission from thin dielectric films bombarded by positive ions.

What is the origin of these dielectric films? There are many reasons for appearing dielectric layers on the metallic surfaces. In the case of gaseous detectors this dielectric films can be formed for example due to the metal oxidation, due to the aging effect, due to the dielectric microinsertions and microparticles (including dust) and in some gas mixtures even due to the formation of thin adsorbed layer of gases or liquids.

Contribution of the aging effect to the spontaneous current growth from the cathode was observed by many authors (see [23]). The conventional explanation of this current growth a single electron emission appearing in some specific place of the detector due to the Malter effect [23].

However, the phenomena we observed are beyond the "classical" Malter effect: we discovered that due to the ion bombardment:

a) $\gamma_+$ (and $\gamma_{ph}$) temporally increase (for 1-30 min)

b) quantum efficiency of the metallic surface to UV and visible photons also temporally increase (for 1-30 min)

c) spontaneous current increase during the ion bombardment may leads to a breakdown (in the case of aging –triggered "classical "Malter effect the current increase, but does not trigger a breakdown or at least this effect was not studied from this angle)

d) Electron emission from the cathode is rather in form of electron jets than single electron emission (see paragraph 4 for more details)

e) After the breakdown one cannot immediately apply the same voltage as it was before the breakdown because new breakdowns appear at considerably lower voltages. So one have to wait for a minutes or more (sometimes for hours!) before the nominal voltage can be applied

f) Adsorbed layer of liquids or some gases also may contribute to the effect described above-see **Fig. 21**. As a result there is probably a possibility to improve rate characteristics of some detectors via the gas optimization

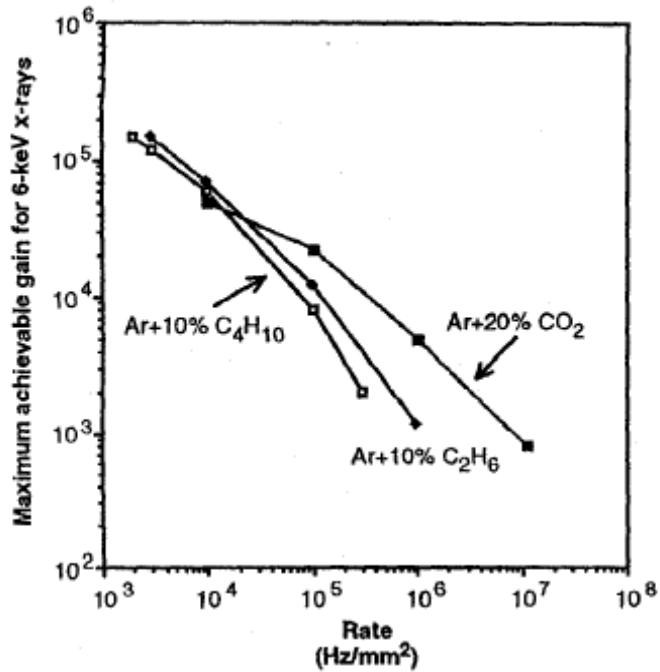

**Fig. 21**. Some preliminary measurements demonstrating that rate characteristics of PPAC may depend on a gas composition

Thus that the cathode excitation effect contributes in operation of many gaseous detectors.

Why the ion current can be suddenly increase and cause the breakdown? One can speculate that this phenomenon occurs on a statistical base: the cathode surface as well as the ion current density j is never perfectly uniform and if at one or two points j is statistically increases, this will cause increase in $\gamma_+$ and $\gamma_{ph}$ and further increase of the j at these point which leads to the local current growth. At some moment the condition for the slow breakdown may be fulfilled:

$$A\gamma=1$$

and this finally leads to a breakdown.

Certainly this effect is very important in detector operation and deserves further studies

## 4.2. CsI photocathodes

Similar to the metallic cathode, it was also observed the after intense ion bombardment the sensitivity of the CsI photocathode (combined either with a single-wire counter or with a MWPC or with a TGEM) to UV and especially to visible light temporally increases (see **Fig. 18, 19**). After the intense ion bombardment is stopped, spurious pulses usually appear for some time (similar as shown in **Fig.20**). A very strong hysteresis affect (see paragraph 3.1) also appears after a discharge in the detector. In some cases after discharges one cannot apply the nominal working voltage to the detector for 20-30 min. These our early observations were recently fully confirmed by COMPASS RICH group [24]. In their detector, after appearing discharges the nominal voltage was not possibly apply sometimes for hours.

Thus the cathode excitation effect can seriously disturb an operation of the CsI –based RICH counters and again it is an additional argument supporting that it is a very important effect deserving detailed studies in the frame of the RD51 collaboration.

### 4.3. Resistive cathodes

The cathode excitation effect was also clearly observed in the case of detectors with resistive electrodes: RPCs and RETGEMs. This effect was systematically studied in the case of high-rate RPCs (an RPC with low resistivity electrode made, for example, of GaAs) developed earlier for us for medical purposes [25]. The same main features were observed in this case as with detectors described above: a hysteresis effect (which was finally minimized by the gas optimization) and afterpulses which are the enemy in high contrast image taking (they can smear the image sharping). As an example, in **Fig. 20** is shown rate dependence of the spurious pulses after the high – rate RPC was irradiated by a strong flux of x-rays. One should consider this curve as an illustration only, because the afterpulses rate depends on the intensity of X-ray flux, the gas gain and the gas composition and the cathode material.

Note that even more pronounced hysteresis effect was sometimes observed in the case melamine or bakelite RPCs: after strong discharges or after strong irradiation (made for aging purposes) one can not apply the nominal voltage for these detectors for days (CMS RPC experience). By the way, bakelite and melomine RPCs always have quite high rate of noise pulses. In the next paragraph we will present some results of noise pulses studied for glass RPCs

Finally we would like to mention our recent studies of CsI coated RETGEM [26]. As in previous cases, if a breakdown happens in the RETGEMs (especially in the case of the CsI coated RETGEMs), one have to wait for 5-10 min before the nominal voltage can be applied. Some spurious pulse may accompany this effect.

To conclude this paragraph: the cathode excitation effect it is a general phenomenon deserving more detailed studies.

### 5. More about jets

Now we came to the most interesting topic of this report - jets triggered breakdown.

As was illustrated by **figures 12** and **14** two effects contribute to the breakdowns at high counting rates:

1) Spontaneous current growth (see **Fig. 14b**), which presumably is due to the cathode excitation effect, and

2) High amplitude pulses (see **Fig 14a**) containing each a great number of electrons emitted during a short time interval from the cathode surface. Let's again stress that the origin of these pulses are beyond a "classical Malter effect' traditionally dealing with single electron emission.

Numerous relatively fresh studied performed in the case of vacuum breakdown gave as important new information about the role of positive ions in seating on the top of dielectric films. According to [27] any, even specially cleaned metallic surface always contains microdielectric insertions. It could be unavoidable insertions between grains, Si microdrops and many other dielectric spots. If positive ions are accumulated on the surface of such microinsertions they create a very strong internal electric field (this fact was basically well known from a Malter era). The new effect, discovered in vacuum breakdown studies, is that this strong electric field causes a slow accumulation in the dielectric film electrons from the metal (due to kind of tunnelling effect) until suddenly a powerful explosive emission of electrons from the film occurs (see **Fig. 22)**. These jets of electron were well observed experimentally in vacuum breakdown studies (see **Fig.23**) and the phenomenon is called "an explosive field emission".

Interesting to note that an "explosive-type electron emission was also earlier observed in the case of some gaseous discharges, for example arc having cold cathodes [28, 29]

It will be logical to assume that exactly the same phenomena exists in the case of gaseous detectors cathode of which are intensively bombarded by positive ions and this is why peaks of current were observed just before the breakdown (see **Figs 12-14).** The new feature which gases "added" to the explosive filed emission is that the positive ions can be accumulated not only on solid dielectric layers, but also on liquid/adsorbed layers which form in some gases [30].

In very first studies the electron jets were observed in the case of PPAC. However, latest studies reveal that this phenomenon exist in many detectors, including GEMs and RPCs. As an example **Figs 24** and **25**shows oscillograms of the current from the GEM irradiated by a strong gamma flux from the cancer treatment facility Racetrack [22] (Karolinska hospital, Stockholm, Sweden). At low gas gains the shape of the current measured from GEMs electrode exactly repeat the shape of the Racetrack current (as it should be). However, with the gradual gain increase the GEM current begun consisting from spikes (see **Fig. 25**) exactly as it was observed before in the case of PPAC.

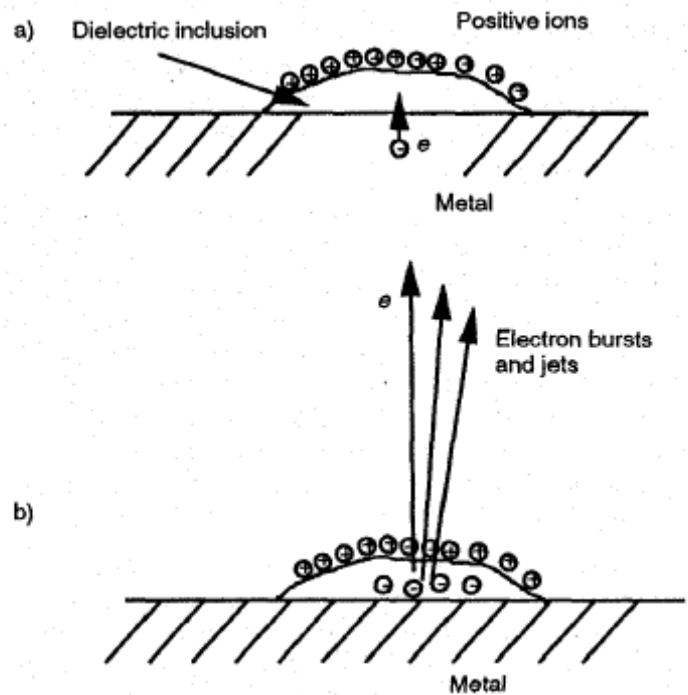

**Fig. 22.** Schematic illustration of a two-step process which leads to emission of jets and bursts from thin dielectric films [16].

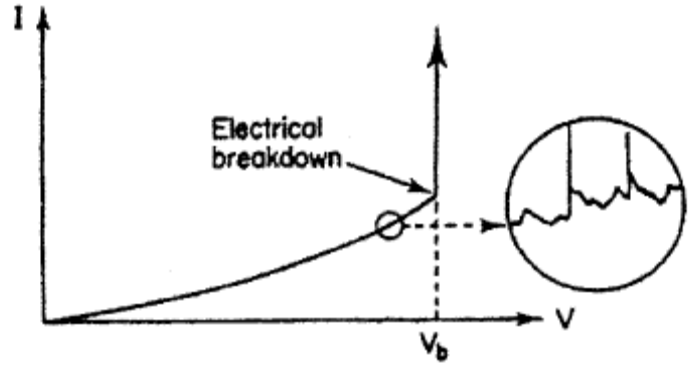

**Fig. 23.** Current-voltage curve in the case of electrical breakdown in vacuum (from [27]). Enlargement shows pulses due to the explosive field emission

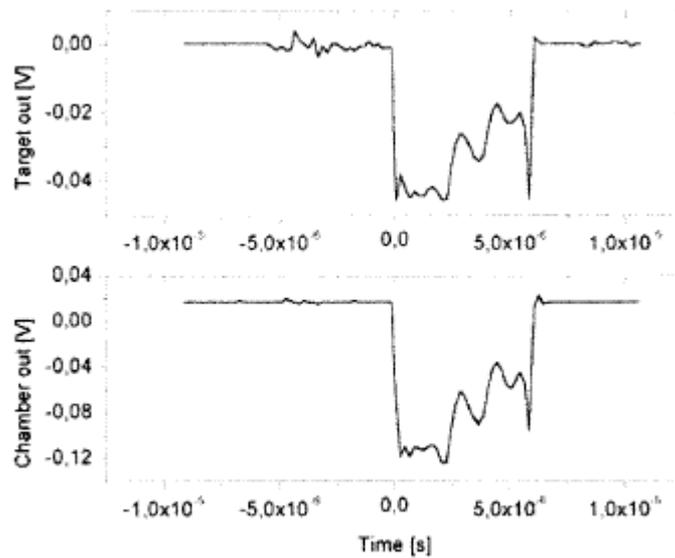

**Fig. 24.** The current from the GEM (at 350 V) recorded directly on a 50 Ω input of the oscilloscope when the GEM was exposed to a pulsed gamma radiation, producing $10^7$ counts/mm on the 2.5 cm x 2.5 cm GEM area. No other resistors (except the 50 Ω input of the scope) were connected. The upper figure shows the current pulse from a racetrack current monitor. The lower figure shows the corresponding current pulse from the GEM readout. The gas mixture Ar+20%CO was used for the measurement (1 atm)[22].

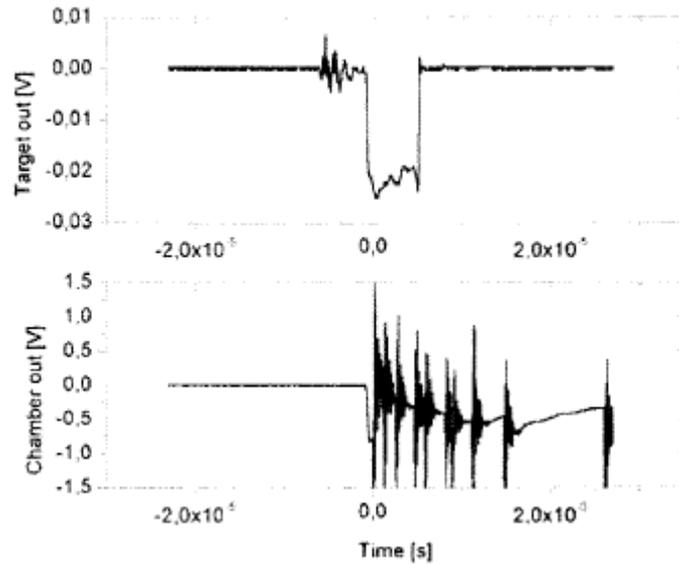

**Fig. 25.** The same setup as in **Fig. 24**, but 420 V applied over the GEM electrodes. The upper oscillogram shows the current pulse from the racetrack current monitor, the lower shows the current from the GEM readout. One can clearly see current pulses of large amplitudes, corresponding to a large number of primary electrons $>>10^5$ [22] .

Interesting conclusions were drown from studies on spurious pulses in RPCs having either a low resistivity cathode -GaAs (a high counting rate RPC) -or high resistivity cathode- glass (low rate RPCs). In **Fig. 26** are shown a pulse height spectrum measured with the High rate RPC operating in proportional mode and detecting: i) single electrons produced from the anode by the UV light (**Fig. 26a**) and ii) spurious pulses (**Fig. 26b**). The spurious pulses were measured at the same voltage without the UV light in anti coincidence with comics. By comparing these two spectrums one can conclude that the mean number of primary electrons which trigger spurious pulses is about 5-10 and thus also can be explained via jets mechanism. This shows that electron jets appear not only during the intense ion bombardment (or in another words during the high rate operation of the detectors), but even much latter (after the bombardment was stopped) in form of afterpulses

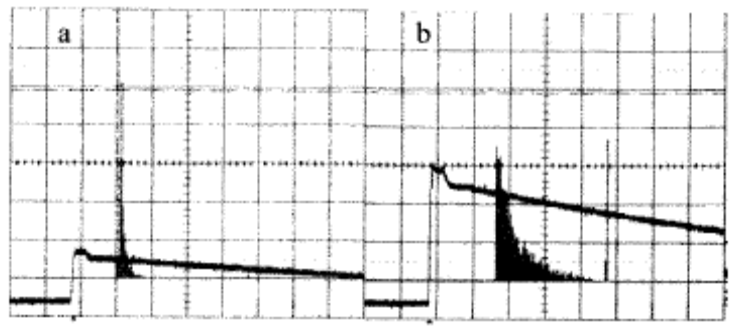

**Fig. 26.** Noise pulses measured in anticoincidence with cosmic: a) Pulse-height spectra of signals from RPCs measured in the case of single primary electrons produced from the cathode by UV emission and (b) in the case of noise pulses. The gas mixture Xe (40%)+Kr (40%)+$CO_2$ (20%) was used (1 atm) [22].

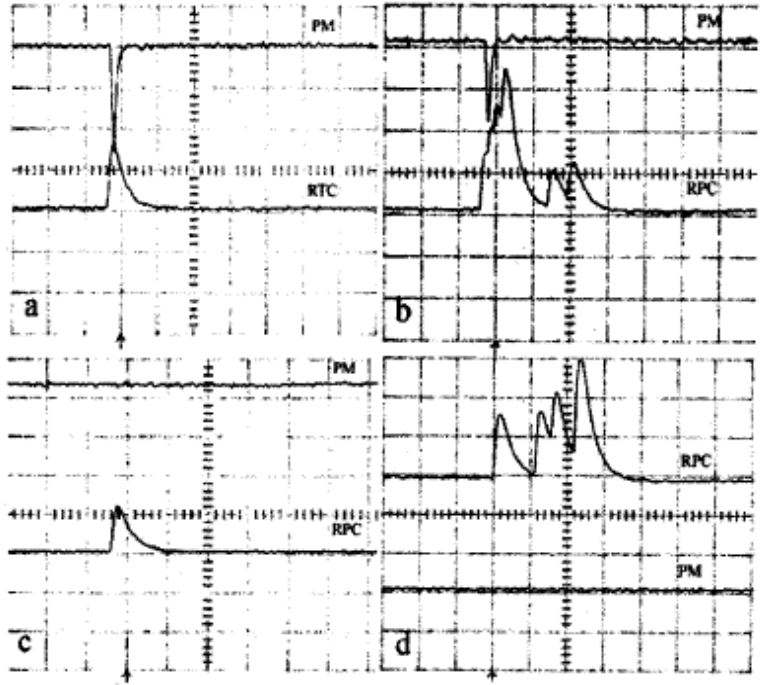

**Fig. 27.** The oscillograms (a) and (b) show signals measured in coincidence with cosmic muons and (c) and (d) noise signals from the RPC [22]. Various voltages were applied to the RPC in the different measurements, (a) and (c) were at V = 7:6 kV and (b) and (d) at V = 8:75 kV. The oscilloscope sensitivity was set to 5 mV/div for the PM signal and 100 mV/div for the RPC. The horizontal scale was set to 0.2 s/div. The gas mixture Ar/Isobutane/Freon (R134) was used in the ratio 48/4/48 [22].

Similar conclusions were drown from measurements of pulses from glass RPCs performed in coincidence and anti coincidence with cosmic muons. As an example **Fig 27a** shows simultaneous pulses from the triggering PM, detecting the light from a plastic scintillator, and from a glass RPC. At a low voltage applied to the RPC the pulse produced by cosmic are narrow and their shape is defined by the ion collection time and by the RC of the equivalent circuit (see **Fig.27a**).

However, with the voltage increase the primary cosmic pulses starts accompanied by afterpulses (see **Fig. 27b**). Their amplitudes and delays with respect to the primary cosmic pulse are randomly distributed within some time interval, duration of which also increases with the voltage. Such behaviour and the time scale do not correspond to a photo [31] - or ion feed back related pulses (which usually have well defined delay times). Thus, again one can attribute their nature to jets. Note also that purely "noise pulses" (which are not in coincidence with cosmics-see **Figs. 27c, d**) also have the same "after pulses" structures at elevated voltages.

## 6. Summary of the results presented in sections 2- 4

Beside breakdowns with well know mechanisms (via the streamers or feedback loops) two new, recently discovered by us mechanisms were described in these chapters.

One of them is the cathode excitation which occurs under intense ion bombardment of metallic cathodes. It was discovered that in this case the coefficients $\gamma_+$ and $\gamma_{ph}$ are temporally increase as well as the quantum efficiency of the cathode indicating that the work function of the cathode was reduced

.

The cathode excitation may case a spontaneous increase of the current cased by the external irradiation and can be also accompanated by electronic jets.

Some authors appeal to the jets breakdown mechanism in attempt to explain some of their experimental data (see for example [32].

## 7. Features of discharge propagation between GEMs and the role of the cathode excitation effect in a so-called "delayed "breakdown

As was already mentioned in paragraph 1.1, with double and triple GEM geometry, due to the enhanced electron diffusion effect in the region of the exits of the holes [6], on can achieve higher values of $Q_{crit}$ than with a single GEM. This is why in most of applications double or triple GEMs are used. However, at some conditions (for example in pure noble gases), another problem may appear: a discharge accidently happened in one of the GEMs may propagate to another one or to the readout plate and this increase the risk for the entire detector and the front-end electronics to be damaged. One should note that nowadays a multistep approach becomes common in MP/SGD s practice: not only GEMs can reliably operate at high gains only if combined with preamplification structures (HERRA experience), but other detectors as well for example capillary plate, TGEMs, RETHEMs, MICROMEGAS.

Thus the discharge propagation in cascade detectors is an important issue.

Discharge propagation between GEMs was studies in several works (see for example [33-36, 38]. Below we will summarize the main results

### 7.1. Setup

The experimental setups for studies of discharge propagation between GEMs and from the GEM to a readout plate are shown in **Figs. 28a and b**. In **Fig. 28a** is shown a setup in which the primary ionization was produced by X-rays from an X-ray gun (18-60keV). In a few words, it is a gas chamber containing double GEM, a drift mesh and the readout plate. In **Fig. 28b** is shown a setup with a single GEM used for studies of discharges propagation from the GEM to the readout plate in which the primary ionization was produced either by alpha particles ($^{241}$Am) or simultaneously by the alpha particles and x-ray photons from the X-ray gun. Both chambers were flushed either with Ar+20%$CO_2$ or with Ar+5%isobutane at a total pressure of 1 atm. With simple capacitor dividers one could simultaneously measure discharge signals from all GEMs electrodes as well as from the drift mesh and the readout plate.

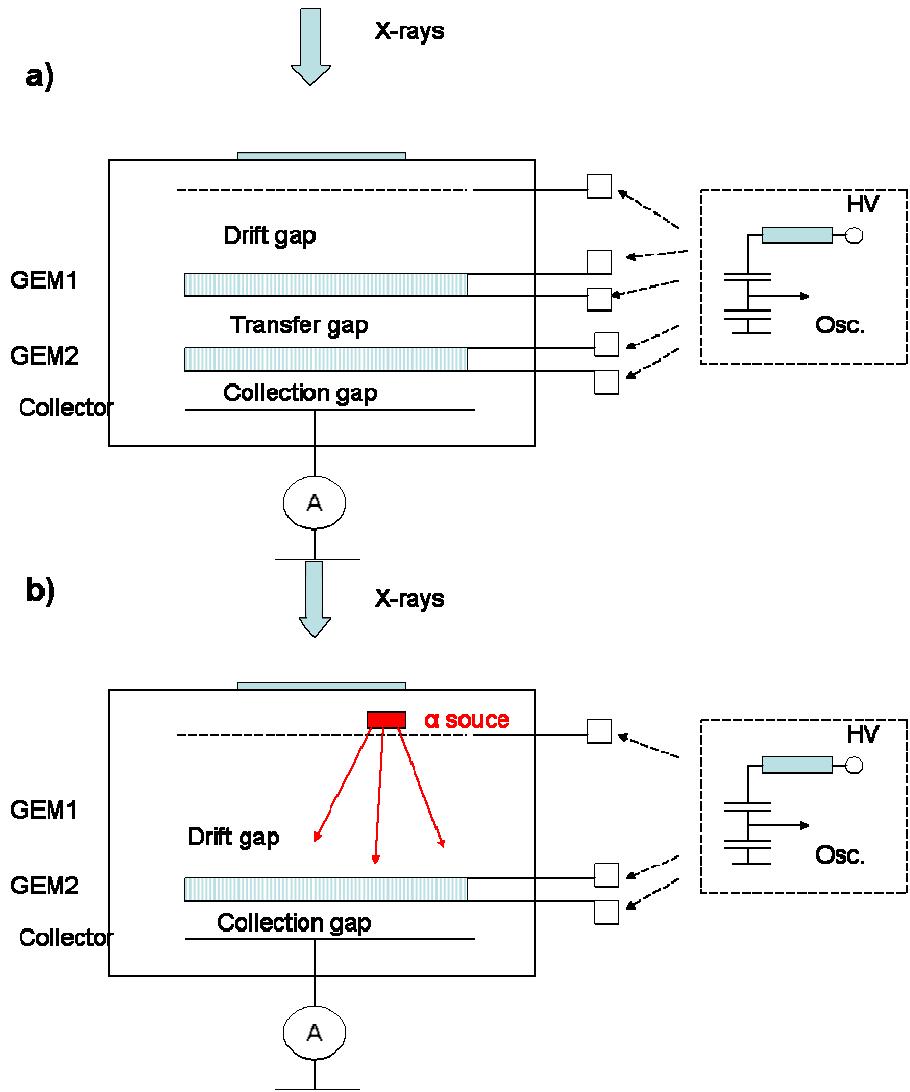

**Fig. 28a**. Schematic drawing of the experimental set up for studies of discharge propagation from one GEM to another GEM and to a readout plate: a) a double GEM detector irradiated with an x-ray gun, **b)** a single GEM detector irradiated by alpha particles or simultaneously with alpha particles and x-rays

## 7.2. Results with X-rays

### *7.2.1. Breakdowns in one GEM*

As an example, in **Fig. 29** are shown oscillogramms of signals from all electrodes of double GEM detector when a breakdown happened only in one GEM, in this particular case in GEM1

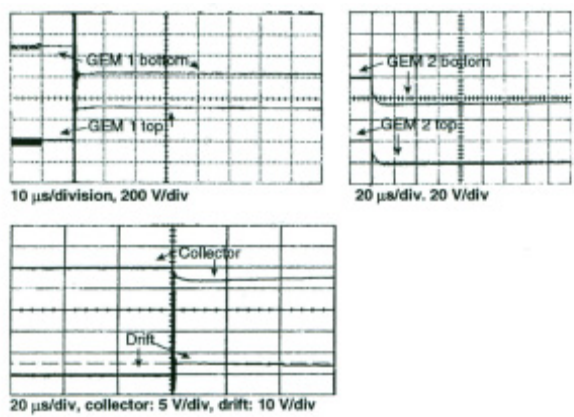

**Fig. 29.** Signals measured simultaneously from all electrodes of the double GEM detector in Ar+20%$CO_2$ gas mixture at p=1atm, when breakdown happened in GEM1

During the breakdown there is short circuit between GEM1 electrodes, and the top and the bottom electrodes come to the same potential and this is why the observed signals have different polarities. Contrary, the pulses seen on the top and the bottom electrodes of GEM2 have the same polarity indicating that they have pick up (induced) origin. The more accurate studies reveal some details. For example, the measurements with a pulses generator demonstrated that the ratio of pick up signal on electrodes of GEM2 to the signal on the GEM1 is 0.063, which was not however observed the case when the breakdown happened.

As one can see from **Fig. 29**, signals from GEM2 are in fact the sum of two signals: a sharp rise signal and a slow rise signal. The sharp part is a true pick up signal due to the capacitive coupling of the GEMs. If the signals from GEM 1 were magnified to the same scale as GEM2, the curved part of the signal could be seen on GEM1 as well -see **Fig. 30**. A probable explanation is that the curved signals are caused by ions moving between the double GEM electrodes-see **Fig. 31**: after the breakdown a bulb of plasma under the GEM forms [37] and ions moving upward induce signals on all electrodes. The size and the shape of the ion signal are sensitive to the potentials on all the electrodes in the chamber as well as on a counting rate. When all electric fields were kept constant except for the field in the transfer gap, the size of the in signal increased with the transfer field.

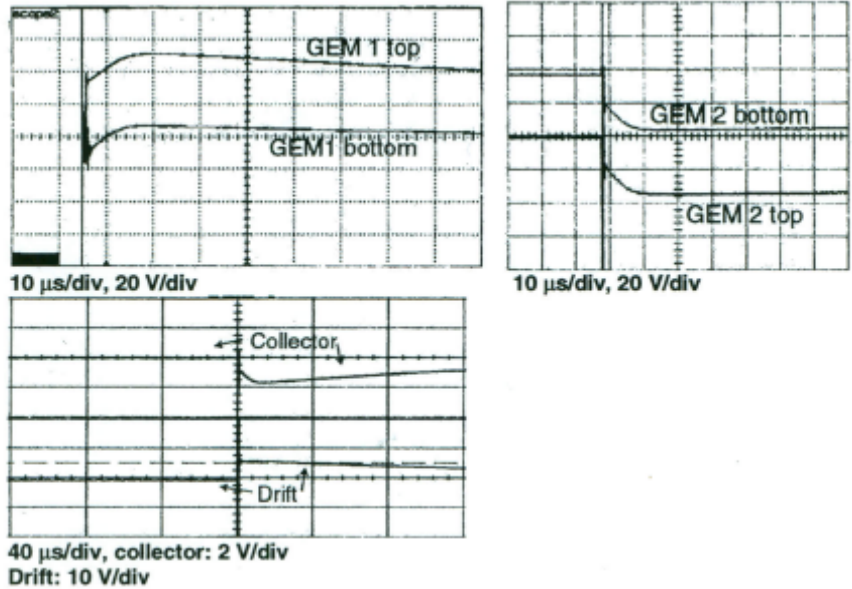

**Fig. 30.** At the increased sensitivity of the scope one can see slow rise signals from GEM electrodes as well. Gas mixture: Ar+20%$CO_2$ at p=1atm

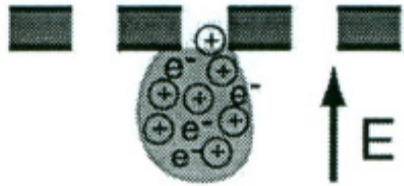

**Fig. 31.** According to F. Fraga [37] et al] a plasma bulb is produce under the GEM hole during the breakdown

### 7.2.2. Discharge propagation from GEM to GEM

When there was propagation from one GEM to another one, the discharge pulses (with a sharp front) were seen on both GEMs (see **Fig. 32**). The signals on GEM2 may become slightly asymmetrical depending on what fraction of discharge charges were collected on the collecting electrode. The gain of GEM2 must be close enough to the breakdown limit to ensure the discharge propagation from GEM1 to GEM2.

Gas gains, counting rate and the distance between GEMs define the entire breakdown limit for double GEM. For GEM1 the limit in the gas gain (i.e. the maximum achievable gain) was fund to be higher than for GEM2. GEM2 was the weakest part of the detector (this is because the total charge in avalanches was highest in this GEM due to preamplification on GEM1). **It was found that breakdown propagation is almost independent on the electric strength between the GEMs. For example, in several occasions the propagation could occur at reversed fields between the GEMs, i.e. a larger negative potential on GEM2 top than on GEM1 bottom. Also, when the distance between the GEMs was small, for example 3mm, a breakdown could propagate upwards, to GEM1 if the discharge was initiated in GEM2. However, this propagation from GEM2 to GEM1 never occurred in the case of large transfer gap, for example 26mm and more.**

These results were latter fully confirmed by Sauli group [38] who measures the discharge propagation probability between two GEMs at various electric fields between them. Their results are presented in **Figs. 33a, b**

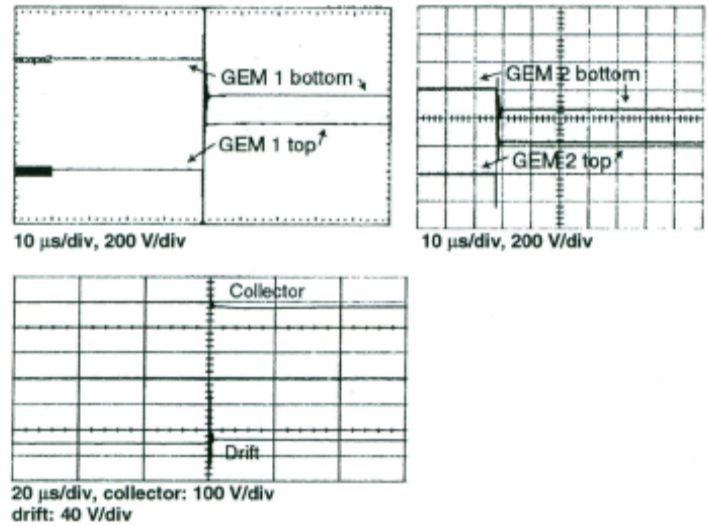

**Fig. 32.** Oscillogramms of signals from all electrodes of the double GEM detector when a breakdown propagated from one GEM to another. In this case the "breakdown" signals were seen on both GEM1 and GEM2. Gas mixture: Ar+20%$CO_2$ at p=1atm

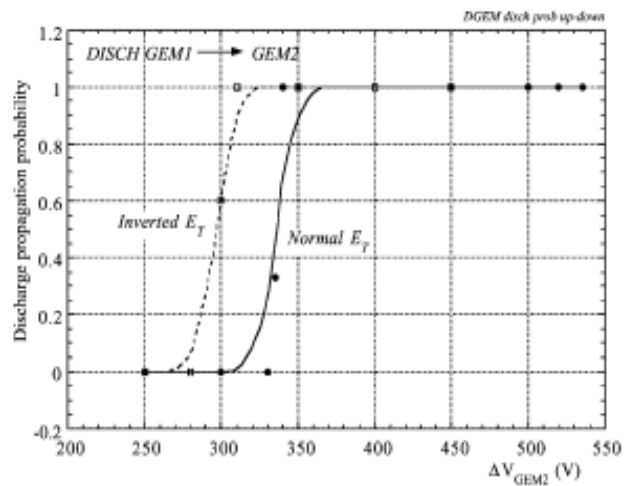

ty between first and
n of voltage on the
fields.

**Fig. 33a.** Discharge propagation probability between first and second GEM as a function of the voltage on the second for normal and inverted transfer field (from [38]).

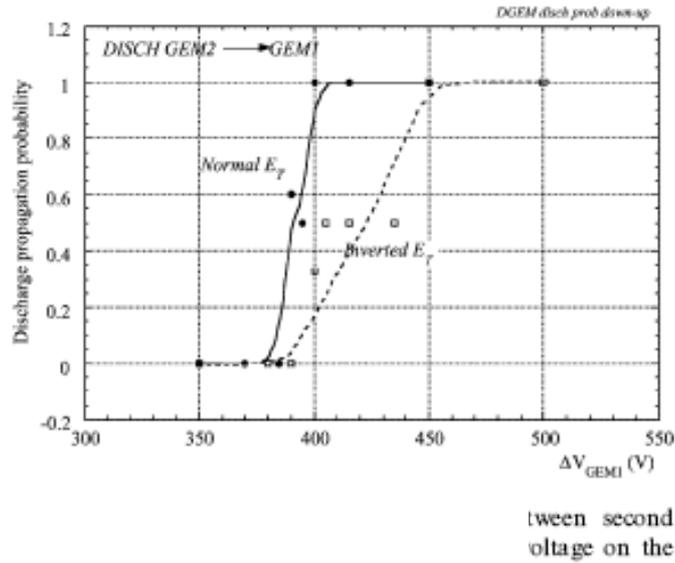

**Fig. 33b**. Discharge propagation probability between the second and the first GEMs as a function of the voltage on the first for normal and inverted electric field (from [38]).

In order to identify the mechanism of the breakdown propagation a time delay between breakdowns in GEM1 and GEM2 was measured. These measurements reveal that with an accuracy of ~10ns there was no time delay between the discharges in GEM1 and GEM2 neither with 3 or 26 mm between GEMs clearly indicating that photon mechanism is responsible for the discharge propagation (see paragraph 7.4a)

### 7.2.3. *Breakdown propagation to the collector*

Under certain circumstances, a breakdown in GEM2 can propagate all the way down to the collector plate. This is the most undesirable scenario in the double GEM detector operation since the readout electronics could be destroyed. Both electrodes of GEM2 drop to ground potential during a breakdown to the collector. Since the applied negative potential before the breakdown on GEM2 top was higher than on the GEM2 bottom, a huge signal was seen on the top electrode and a smaller one on the bottom electrode-see **Fig. 34**. The potential of both electrodes goes down and thus the signal jumps both have the same polarity. At the same time a very large current pulse was seen on the collector.

The condition for the breakdown propagation to the collector was that the electric field strength between the GEM2 bottom and the ground had to be above a critical value, ~10kV/cm. This was confirmed by Sauli group (see **Fig. 35** and **36** from [38]).

When the field strength was lower the critical value, the discharge remained confined in GEM2 holes and did not spread down to the collector plate

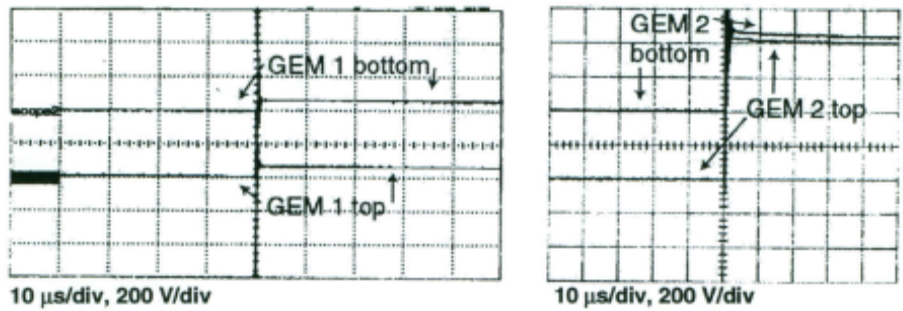

**Fig. 34**. Oscillograms of signals when a breakdown in GEM2 that propagates down to the collector plate. On GEM1 pick up signals are seen. The signal from GEM2 top was large and had the same polarity as the signal from the GEM2 bottom

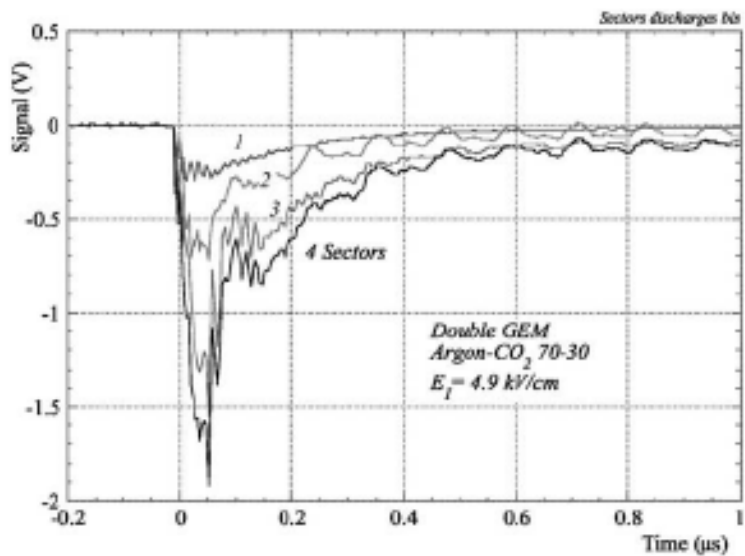

**Fig. 35.** Discharge signal on anode for increasing GEM capacitance obtained by grouping one to four sections [38].

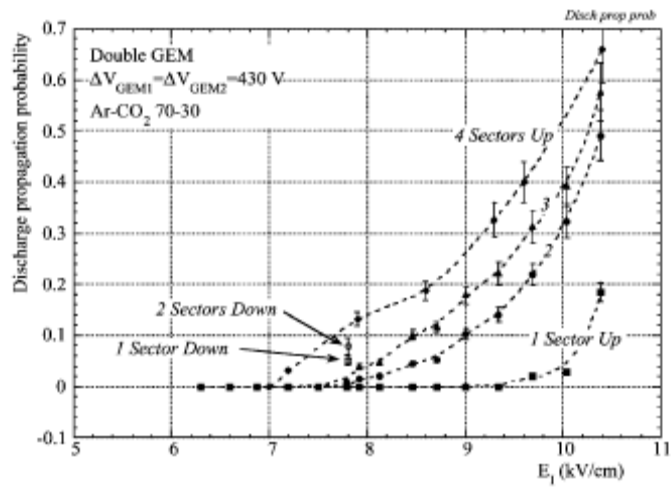

**Fig. 36** Discharge propagation probability as a function of induction field for a sectored GEM(from [38]). Note that effect of the capacitance on the discharge propagation was firs observed by HERRA group and then was also studied by a Swedish group (see [36]).

This group also noticed that the discharge probability depends on energy of sparks which was increase by adding capacitors (see **Fig. 35** and **36**)

## 7.3. Results with alpha particles

### 7.3.1. *Discharge propagation to the collector*

As was mentioned earlier, sometimes during the breakdown the anode and the cathode of GEM2 did not share the voltage applied over the GEM evenly as shown in **Fig. 32**. This was due to the partial discharge propagation to the collector. With X-rays the discharge could propagate from GEM2 to the collector at a critical electric field $E>E_{crit}$~10kV/cm. In contrast, in the presence of alpha particles the discharge <u>always</u> fully or partially propagated to the readopt plate. The fraction of the discharge charge collecting on the readout plate (we called it semi-propagation) gradually increased with the electric field between the GEM bottom electrode and the readout plate. This is illustrated by **Fig. 37** showing oscillograms from all detector electrodes at three gradually increased fields between the GEM bottom and the collector. As one can see the current on the collector also gradually increase.

Only at high enough filed there is a full propagation to the collector during which the potential on both GEM electrodes went to ground-see **Fig. 35**.

Indirectly these results were confirmed by Bachmann et al [38]. These authors did not study partial propagation and simply measured the probability of the propagation.

In studied of full propagation to the collector electrode a large capacitor, 0.1 µF, was connected between the collector and the ground. This made it possible to see the amount of charge going down to the collector during the discharge. A maximum of 400nC ($10^{12}$ electrons) went down from the GEM to the collector during a full propagation of the GEM breakdown

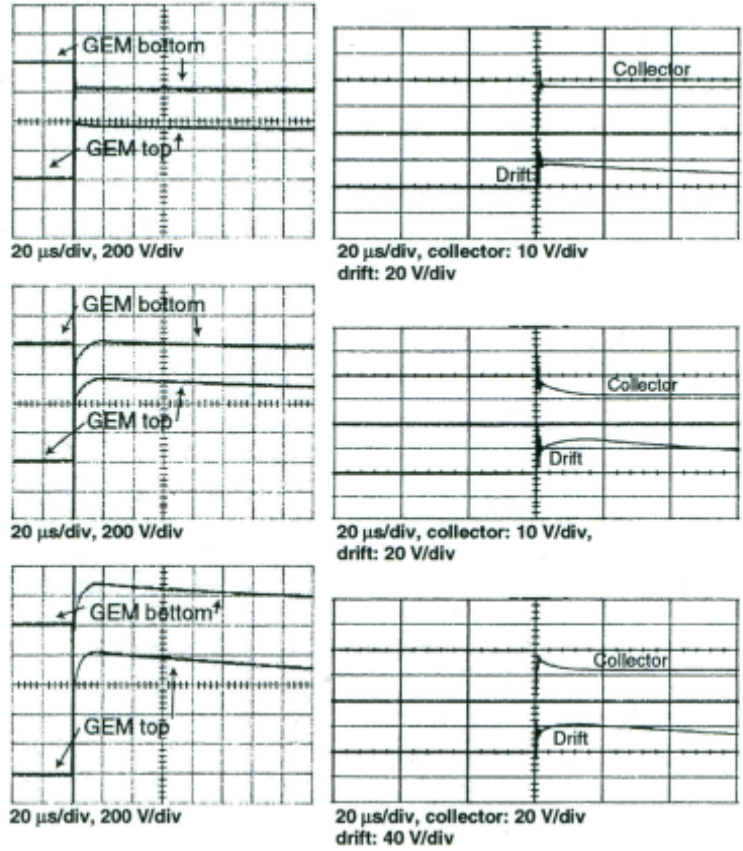

**Fig. 37.** Oscillograms of signals from all detector electrodes illustration a semi-propagation of the discharge from the GEM to the collector

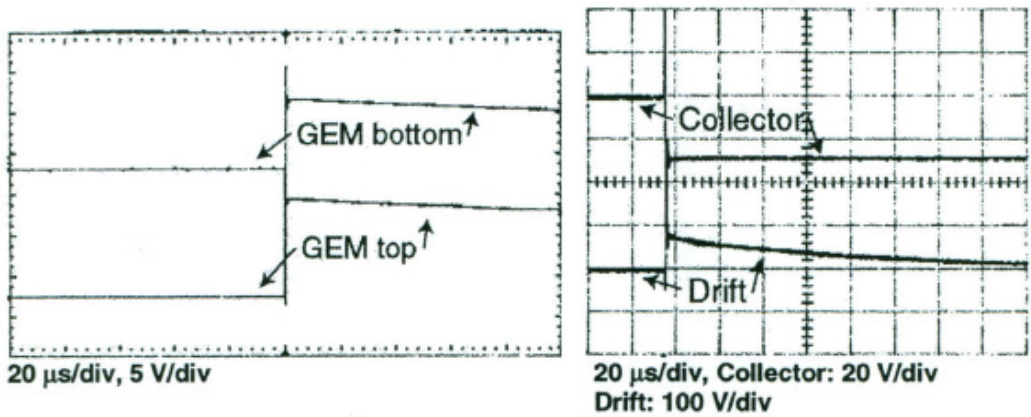

**Fig. 38.** Oscillograms of signals during the fill propagation of the discharge from the GEM to the collector

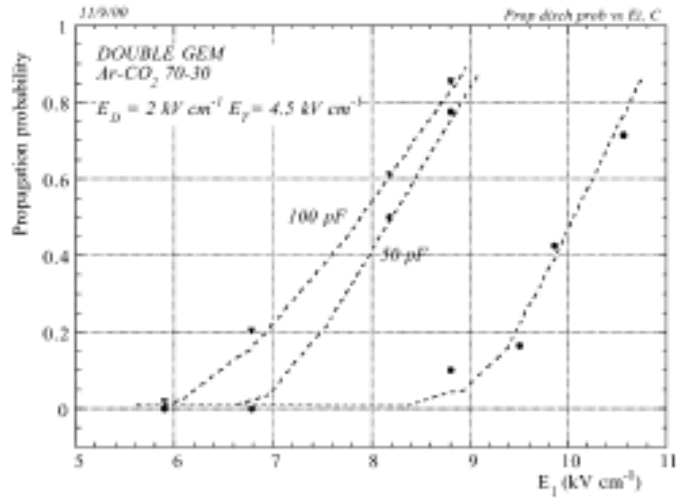

Fig.39. Discharge propagation probability without a (right curve) and with (left curves) additional capacitors in parallel with the induction gap [38].

### 7.3.2. Breakdowns with a time delay between them

An interesting phenomenon was observed with a large area GEMs ($40 \times 40 cm^2$) when two consequent breakdowns may appear at high collection fields. To study this effect on smaller detector, GEMs electrodes were connected to the ground via 5nF capacitors as shown in **Fig. 40**; in this case two consequent breakdowns appeared at high collection fields.

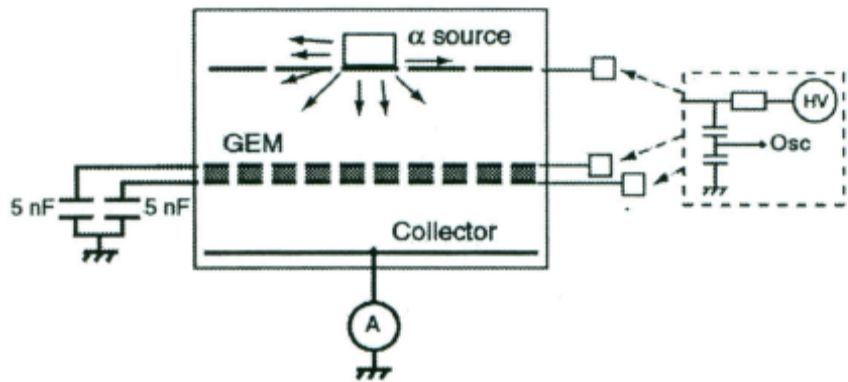

**Fig. 40.** A setup for studies of breakdown propagation when GEM electrodes were connected to ground via 5nF capacitors

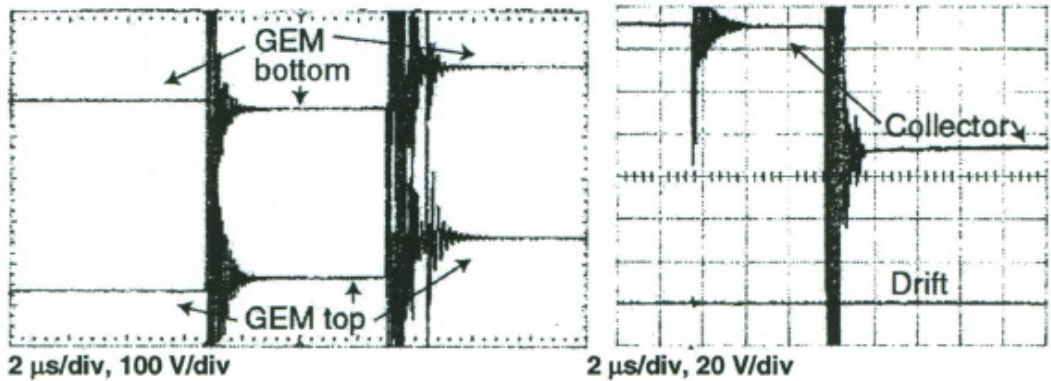

**Fig. 41**. Two breakdowns following each other: the breakdown in the GEM was followed with some delay by discharge propagation to the collector

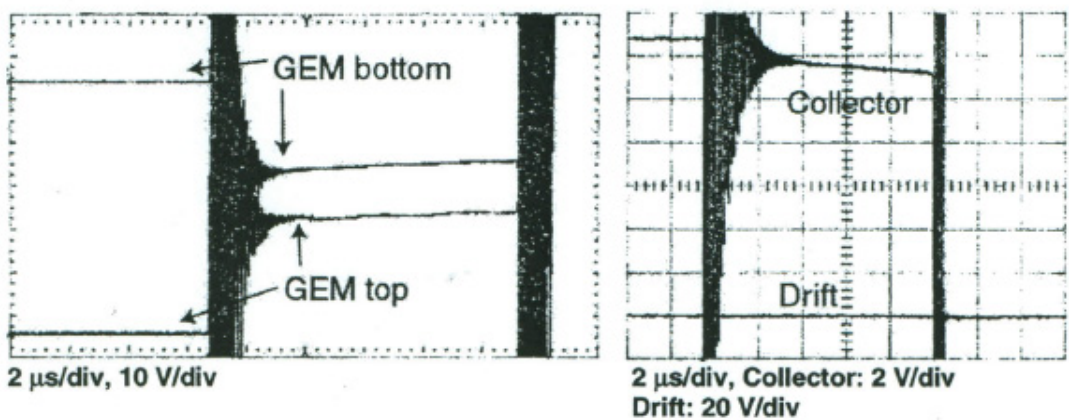

**Fig. 42.** In a more sensitive scale one can see a steady current increase before the second breakdown happens

First there was a discharge in the GEM and with a time delay the breakdown propagated down to the collector-see **Figs. 41, 42**. In order to reduce the energy content during a discharge down to the ground, the distance between the GEM and the collector was decreased from 3mm to 1mm. The time delay between the consequent breakdowns varied very much, from 15 to 25 μs. Between the breakdowns a positive current slope was seen on both GEM's electrodes and a negative slope on the collector-see **Fig. 42**. This phenomenon very much remind a current grows observed in the case of the cathode excitation effect (see **Fig. 12**)

### 7.4. Interpretation of the results

#### 7.4.1. Discharge propagation from GEM to GEM

The fact that there was practically no delay between breakdowns in two GEM (with accuracy of 10ns) proves that neither electrons nor ions can be responsible for the breakdown (their drift time for a 3mm gap are 50-70 ns and 60-130 μs correspondingly. Thus the discharge propagation was performed by avalanche photons. What is the exact mechanism of this propagation? The possibility that the discharge from the GEM2 propagate upwards by a steamer mechanism can be ruled out because the propagation occurs in very weak fields and even at the reverse electric filed in the region between the GEMs (for the streamer propagation much stronger electric field is necessary [1].Thus one can assume that the main mechanism is a creation photoelectrons in the drift region which in turn triggers a discharge in the GEM1 if the Raether limit is satisfied. Schematically this is illustrated by **Fig. 43.**

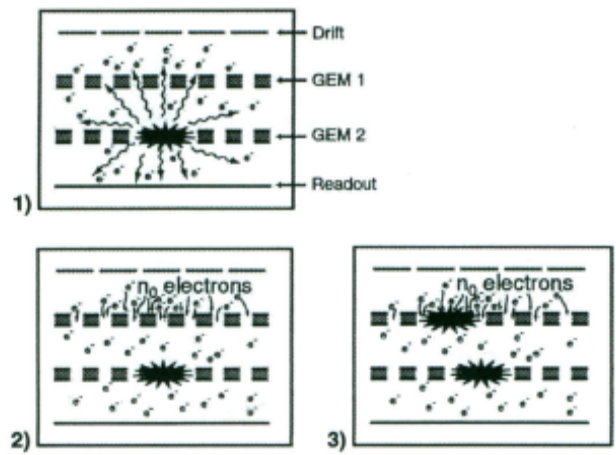

**Fig. 43.** A schematic drawing illustrating of a possible mechanism of discharge propagation from GEM2 to GEM1. The UV photons from the discharge in the GEM2 photoionize gas in the entire detector, including the drift region. The secondary electrons produced there trigger a breakdown in GEM1 (if the Raether limit is satisfied)

### 7.4.2. Breakdown propagation to the collector with X-rays

From the practical point of view the most important phenomena is discharge propagation to the readout plate. With X-rays breakdown propagation down to the collector was observed at electric fields above 10V/cm in the gap. It is known that discharge in the GEM creates a plasma bulb bulb below the holes which in the strong field may triggers streamers via photon mechanism-see **Fig. 44.**

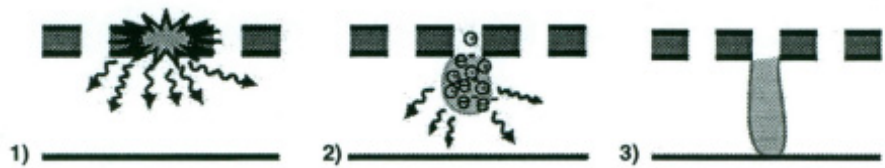

**Fig. 44.** A possible qualitative explanation of the breakdown propagation from the GEM to the collector with X-rays. According to this model the breakdown produce photons and a dense cloud of electrons and ions under the GEM and this in $E > E_{crit}$ creates a streamer

### 7.4.3. Semi-propagation and full breakdown propagations to the collector with alpha particles

Alpha particles produce heavily ionized tracks and when there is a discharge in the GEM these conductive passes make it possible for the electrons to move down to the collector: the electrons move through preionized media-tracks. As a results at almost any field between the bottom GEM electrode and the collector part of the breakdown charge transfers to the collector as schematically shown in **Fig. 45.**

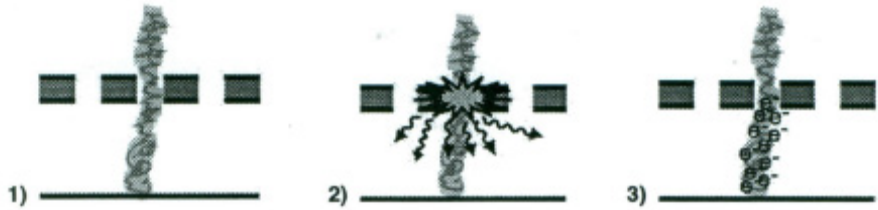

**Fig. 45.** An illustration of the hypothesis of the semi-propagation of a discharge from the GEM to the collector with alpha particles. Alpha particles moving through the gas in the detector create dense ionised tracks. When the discharge appears, the electrons can easily move down to the collector through this preionized channel and cause a discharge: all electrostatic energy stored in the GEM capacity can be released via the conductive pass

### 7.4.4. Breakdowns with a time delay

The delayed breakdown can be explained by a cathode excitation effect. After the spark in the GEM the GEM's cathode region close to the spark gets "excited" due to the intense ion bombardment during the spark. The ions from alpha particles that are collecting during a few μs after the initial breakdown, will then bombard the already "excited" surface-see **Fig. 46**. Due to the lowering of the cathode work function the condition for the feedback loop $A\gamma_+ =1$ can be satisfied at some moment even at a very low gas gain in the region between the GEM and the collector. Experimentally it appears as a slow current growth. Breakdown then appears due to ion feedback or, more likely, by a combination of the ion feedback with the electron jet emission

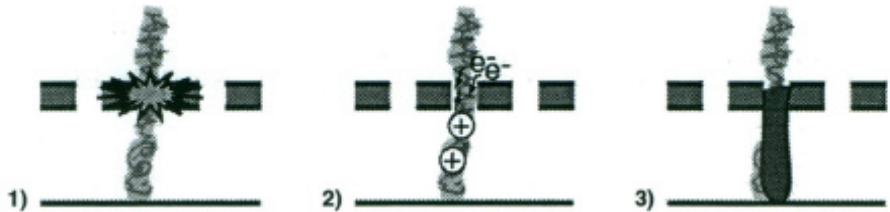

**Fig. 46.** A schematic illustration of the hypothesis pretending to explain delayed breakdown. When there is a spark in GEM triggered by alpha particles, the cathode will emit for some time electrons due to the slow collected ions from the alpha track via a secondary electron emission from the cathode due to the ion recombination there. This may cause another breakdown in the space between the GEM and the collector due to the combination of two effects: ion feedback and jets

### 7.5. A short summary of the discharge propagation studies:

Since no time delay was found for the GEM to GEM propagation and since it was independent of the electric field between the GEMs it was concluded to be due to photomechanism, i.e. photons propagate the discharge. The hypothesis is that ultraviolet photons created in the breakdown in a GEM ionize gas molecules in the detector and the created photoelectrons are injected into the other GEM and cause a breakdown.

Possible ways to suppress propagation of discharges between GEMs was found to be anincreased distance between GEMs or a reduction of the gain in the ''receiving'' GEM.

Ways to suppress breakdown propagation down to the readout could be to increase the distance between the GEM and readout and to keep the electric field below the critical value of~.10 kV/cm

One possible explanation to breakdowns with delay could be the so-called ''cathode excitation effect''. There is no doubt that after the spark in the GEM the cathode region close to the spark gets ''excited''. The ions from alpha particles that are collecting during a few ms after the initial breakdown, will then bombard the already ''excited'' surface. Due to the lowering of the work function the condition for the efficient feedback loop $A\gamma_+=1$ can be satisfied at very low gas gain in the region between the GEM and the collector.

Experimentally, it appears as a slow current growth. Breakdown then appears due to ion feedback or, more likely, by a combination of the ion feedback with the jet emission.

## 8. Feature of discharges in MP/SGD s with resistive protective layers

Besides streamers, feedbacks and the cathode excitation effect another phenomena was observed in some MP/SGD s at certain conditions– an appearance of a glow discharge. In MP/SGD s with metallic electrodes the glow discharge may appear for example when they operate in pure He and Ne at reduced pressures. Let's however concentrate on most important practical case -on the glow discharges which appear in detectors with resistive electrodes operating at normal pressures and in quenched gases.

Resistive electrodes were implemented in some MP/SGD s designs recently to restrict the energy of discharges [26,39]. During the last year this list of detectors was enlarged by MICROMEGAS [40, 41]

Unfortunately, resistive protection layers may restrict the rate capability of the detectors. However, as was shown in the case of the large gap (2-3mm) RPCs [39] and small gap RPCs (0.1-0.4mm) [42] the resistivity of the electrodes can be optimized in order to achieve at the same time high the rate capability and preserve their protection properties. As an example in **Fig. 47** are presented gain curves of the large gap (3mm) RPCs as a function of the counting rate for various resistivities of the electrodes. One can see that at some reduced values of the resistivity and reduced gas gain the "metallic rate limit "(see paragraph 2.1 and **Fig. 8)** can be reached. Bases on these studies were successfully developed discharge-protected high rate RPCs for medical applications [43]. These RPCs had a cathode plane made of low resistivity materials Si ($10^{-2}$ -$10^2$ $\Omega$cm) or GaAs ($10^3$-$10^8$ $\Omega$cm) or resistive glass ($10^8$-$10^{12}$ $\Omega$cm) and the glass anode plane coated with metallic strips 30-50 μm pitch; the gap between the anode and the cathode planes was 0.1-0.3 mm. These detectors were use for obtaining mammographic images and with low resistivity cathode (< $10^8$ $\Omega$cm) could operate at counting rates more than $10^5$ Hz per strips at a gas gain of $10^4$.

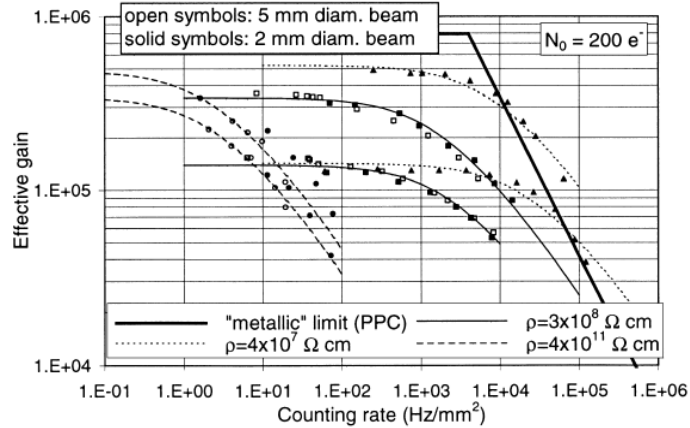

**Fig. 47.** Gain-rate characteristics of the detector for several values of the anode plate resistivity and X-ray beam diameters of 2 and 5 mm. For the lower resistivity studied counting rates of $10^5$ Hz/mm$^2$ were achieved at gains between $10^4$ and $10^5$. The thick solid line marks the intrinsic counting rate limitations for metallic PPAC (see (**Fig.8**)) (from [39]).

During the tests of various cathodes it was observed that in some range of electrodes' resistivity (see **Fig. 48**) and at sufficiently high gas gains (which are below or equal the" metallic limit") a glow discharge appears instead of usual sparks. The resistivity at which the detector transits to the glow discharge is typically in the range of $10^4$-$10^8$ $\Omega$cm. Of course, the boundaries indicated schematically in **Fig. 48** are not very precise and depend on the detector geometry and the gas.

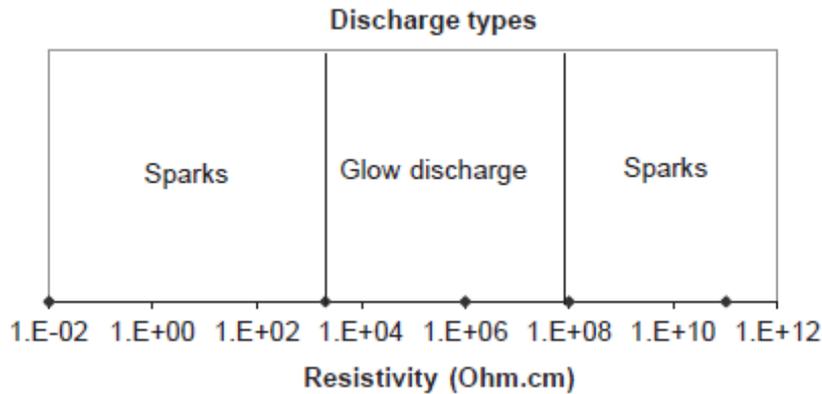

**Fig. 48.** Types of discharges in RPCs with various electrode resistivities (from [44]).

A very similar behaviour was observed in the case of the RETGEMs: at some medium range of their electrodes resistivity ($10^5$-$10^7$ $\Omega/\square$) and at gains closed to the Raether limit a glow discharge appears. As an example, Fig. **49** shows a photograph of the glow discharge which appeared in the Kapton RETGEM (resistivity of~$10^6$ $\Omega/\square$)) operating in Ar

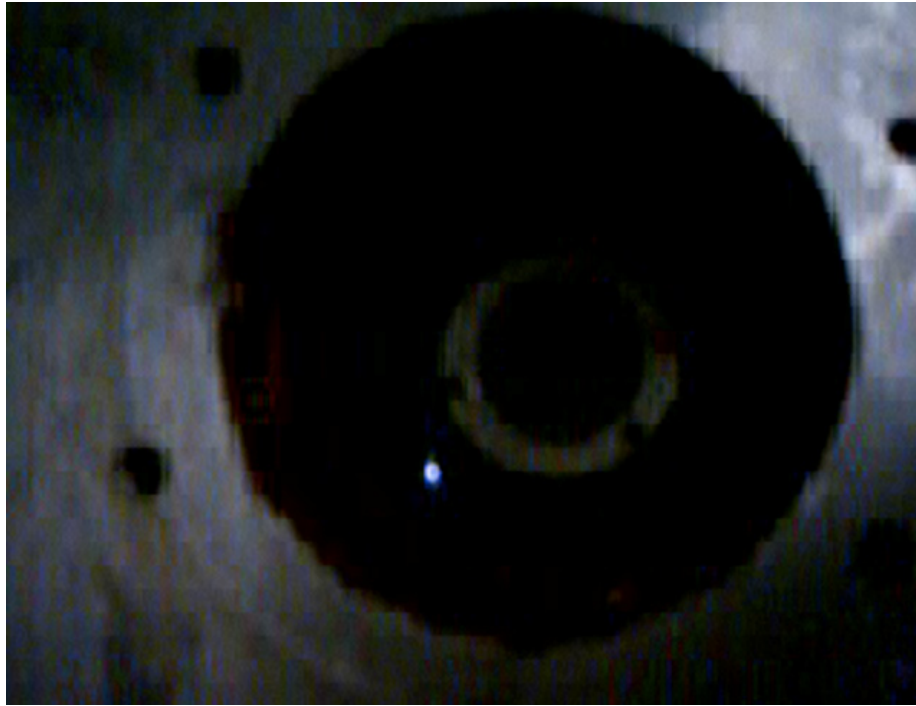

**Fig. 49.** A photo of a glow discharge occurring between two RETGEMS at overall gain >$10^6$. Interesting to note that after~10 min of this discharge the detector continue to operate normally (after the discharge was terminated) –this discharge did not harm ether the detector or the front end electronics (from [45])

## 9. Surface streamers

Finally let's consider another type of discharges in MP/SGD s-surface streamers and surface discharges which often restrict the maximum voltage which can be applied in some detectors, for example MSGCs or GEMs, or TGEM. In this report we will shortly summarize what is known about these "invisible" enemies of gaseous detectors.

Very often in gaseous detectors designs the electrodes are separated by a flat dielectric surface as shown for example in **Fig. 50.** Usually this is the weakest part of the detector from the point of view of the quality the HV holding. If a HV is applied to one of the electrodes whereas the other one is connected to an amplifier than at some voltages V>$V_{crit}$ one can observe spurious pulses from the amplifier. With the further voltage increase the amplitude of these pulses increase and they suddenly may transit to surface streamers and then to a powerful breakdown along the surface. As an example in **Fig .51** are shown surface streamer pulses detected in MSGC with a glass substrate. It was observed that $V_{crit}$ has its maximum value in the case of very clean dielectric surfaces; in this case there is a correlation between $V_{crit}$ and the threshold voltage $V_{tr}$ on the detector when the gas amplification starts: $V_{crit}=kV_{tr}$, where k is a coefficient . This is an indication of the avalanche nature of this process. As was shown in [46] the surface avalanches and streamers on clean surface can easily propagate along the surface on a quiet large distance even in very weak electric fields. This is because these plasma filaments create a strong electric field in the vicinity of their heads (see **Fig. 52**) and also because the photoelectrons necessary for their propagation can be crated from the dielectric surface with lower energetic threshold compared to the gas photoionization.

As was shown in work [47] surface streamers play the major role in breakdown in MSGCs with very clean substrates. Surface streamers may also trigger a breakdown propagating perpendicular to strips (of course in this case a combination of several effects is involved, one of them is that the during the breakdown the closes strips went to the same potential as it happens in cascaded GEMs).

In most of practical cases however, the surfaces are not clean and are covered by various thin semiconductive layers affecting the surface resistivity, for example an adsorbed layer of water, various dirts and so on. In this case with the increase the voltage between the electrodes (see **Fig. 50**) the rate and the amplitudes of the spurious pulses increase and they can gradually transfer to a leakage current.

The maximum voltage at which the spurious pulses appear in the case of imperfect surface $V_{inper} \ll V_{crit}$.

Surface spurious pulses and discharges prevent one in some cases applying the necessary working voltage to the gaseous detectors. For example, in early designs of MWPC gains only $10^3$-$10^4$ were achieve due to the surface problems in the dielectric interface between the anode and the cathode wire planes. Latter by improving the design of this interface it was possible to reach gain up to $10^6$ so the MWPC becomes sensitive to single photoelectrons. A very similar situation is/was with the GEM and TGEM detectors. The maximum HV to these detectors can be applied only if thy are very clean and free from micro particles.

What are the ways of suppressing surface related pulses and discharges? The easiest one is just to create rectangular grooves between the electrodes as shown in **Fig. 50b**. On vertical surface of these groove the electric filed lines are perpendicular to the surface and this very efficiently prevents any leakage charge propagates along this surface.

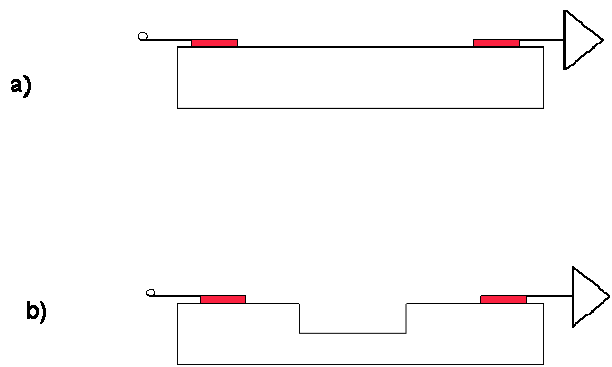

**Fig. 50.** Possible designs of a dielectric interface between the anode and cathode electrodes: a flat dielectric surface between strip electrodes, b) a dielectric surface with a groove to prevent surface streamers

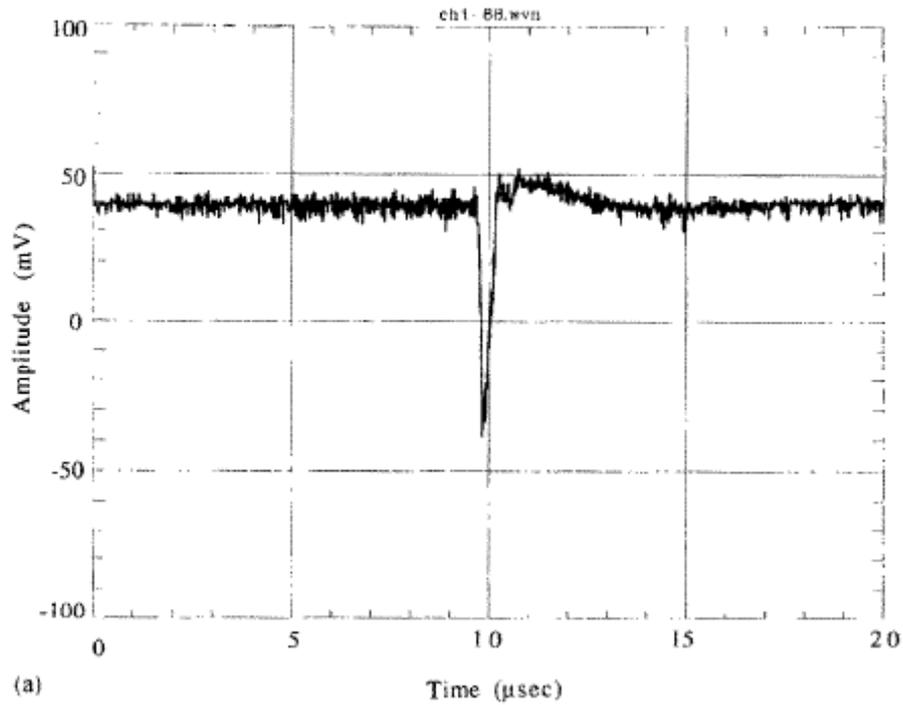

**Fig. 51.** A typical streamer current pulse from the MSGC which appears at voltages close to breakdown (from [47])

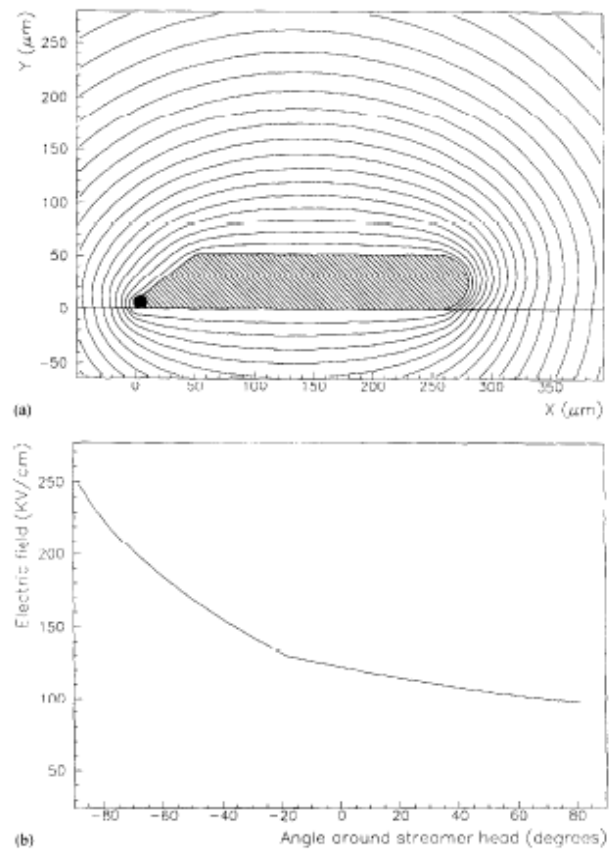

**Fig. 52.** The field strength around the tip of the streamer. The inset details a map of equipotential lines for a streamer near the anode of an "MSGC" in the presence of a surface [47].

## 10. Conclusions

This report describes the latest experimental results on breakdown in MP/SGDs as well as some hypotheses explaining these results. Besides the classical breakdown mechanisms occurring via streamers or a feedback loops new breakdown mechanisms were also identified and shortly described: a cathode excitation effect and electron jet emitting under ion bombardment from the detector's cathodes. It was also shown that MP/SGDs have a clear gain vs. rate limit which one should take in to account when considering high rate applications of MP/SGD such as in LHC or in a medical one.

In spite the fact that the main physical picture of breakdowns in MP/SGD s is qualitative rather clear, there are still many important "details" to be understood. We believe that this report will stimulate other researches to make further studied of these very interesting phenomena

## 11. Acknowledgement

This report was written in the framework of the RD51 collaboration.